\documentclass[showpacs, twocolumn, aps, prb, floatfix]{revtex4-1} 

        \expandafter\let\csname equation*\endcsname\relax
        \expandafter\let\csname endequation*\endcsname\relax
        \usepackage{amsmath,bm}
        \usepackage{mathrsfs}
        \usepackage{amssymb}
        \usepackage{amsfonts} 
        \usepackage{graphicx}
        \usepackage{url}
        \usepackage{hyperref}
        \usepackage[columnwise]{lineno}  
        \usepackage{color}
        \usepackage{epstopdf}  
        
        \newcommand{\V}{\mathscr{V} }
        \newcommand{\deltainf}{\Delta(\infty) }
        
            \makeatletter
            \newsavebox{\@brx}
            \newcommand{\llangle}[1][]{\savebox{\@brx}{\(\m@th{#1\langle}\)}%
              \mathopen{\copy\@brx\kern-0.5\wd\@brx\usebox{\@brx}}}
            \newcommand{\rrangle}[1][]{\savebox{\@brx}{\(\m@th{#1\rangle}\)}%
              \mathclose{\copy\@brx\kern-0.5\wd\@brx\usebox{\@brx}}}
            \makeatother

\newcommand{\E}{{\mathcal{E}} }

\newcommand{\LL}{Luttinger liquid }

\newcommand{\WC}{Wigner crystal }

\newcommand{\Valpha} {$V \textrm{-}\alpha$ }
\newcommand{\dmrg}{density matrix renormalization group }
\newcommand{\mps}{matrix product state }
\newcommand{\mpss}{matrix product states}

\newcommand{\tvmodel}{$t\textrm{-}V$ model}
\newcommand{\SG}{sine-Gordon }

\usepackage{bibentry}

\newcommand{\ignore}[1]{}
\newcommand{\bibfnamefont}[1]{#1}
\newcommand{\bibnamefont}[1]{#1}

\mathchardef\mhyphen="2D

\begin{document}


        \title{Ground states of long-range interacting fermions in one spatial dimension}
        
    \author{Zhi-Hua Li}
    \address{School of Science, Xi'an Technological University, Xi'an 710021, China}

    \begin{abstract}
        We explore systematically the ground state properties of one dimensional
        fermions with long-range interactions decaying in a power law
        $\sim1/r^\alpha$  through the density matrix renormalization group
        algorithm.  By comparing values of Luttinger liquid parameters precisely
        measured in two different ways, we show convincing evidence that
         Luttinger liquid theory is valid if $\alpha$ is larger than some
        threshold, otherwise the theory breakdown. 
        Combining analysis on structure factor, charge gap and charge stiffness,
        we determine how the metal-insulator transition point develops as the
        interaction range is continuously tuned. A region in the range of
        $0\leq\alpha\leq1$ and small interactions has finite charge gaps,
        but, interestingly, it shows metallic nature at the same time.  From
        these, we obtain approximate phase diagrams for the entire parameter
        space and for band fillings equal 1/2 and 1/3.  Finally, we compare
        certain bosonization and field theory formula with our quasi-exact
        numerical results, from which disagreements are found. 
        \end{abstract} 


    \maketitle
    \modulolinenumbers[5]

\section{introduction}
    One-dimensional (1D) electron models are widely studied due to the viability
    of various analytical and numerical methods. Typical examples are the Hubbard
    model and its extensions with short-range (SR) interactions. 
    A dominant theoretical framework for solving and understanding such
    models is the \LL theory
    \cite{haldane1981luttinger,voit1995one,giamarchi2004quantum}.  
    In that theory correlation functions decay as
    non-universal power-laws, with the exponents determined by a single
    parameter $K$ (per degree of freedom). 
    However, experiments on quasi-1D systems such as quantum
    wires\cite{goni1991one,steinberg2006localization} and carbon nanotubes
    \cite{deshpande2008one} highlight the importance of long-range (LR)
    interactions, as in 1D screening is absent and the electrons
    should interact via the bare Coulomb force. Then the Luttinger liquid theory
    is not guaranteed to be effective, since it assumes SR interactions. 

    Schulz \cite{schulz1993wigner} applied bosonization method to 1D
    electron systems with Coulomb force. He found $4k_F$ density correlation
    decaying slower than any power laws, thus beyond Luttinger liquid, and
    interpreted the ground state as a quasi-Wigner crystal. 
    The occurrence of $4k_F$ (or $2k_F$ for spinless fermions) wave vector was
    later examined numerically in continuum  \cite{casula2006ground,
    shulenburger2008correlation, astrakharchik2011exact, ferre2015phase} and
    discrete models \cite{fano1999unscreened,hohenadler2012interaction}.  
    
    The systems studied in
    \onlinecite{schulz1993wigner,casula2006ground, 
    shulenburger2008correlation, astrakharchik2011exact, ferre2015phase} are
    electrons moving in a continuum (e.g. for electron gas confined in
    semiconductor heterostructures);  when moving on a lattice,
    they can be localized and form an charge-density-wave (CDW)
    insulator. The impact of LR interaction on the metal-insulator
    transition was studied in
    \onlinecite{poilblanc1997insulator,capponi2000effects} using exact
    diagonalization, and it was shown that the umklapp scattering is reduced
    in the presence of LR interaction.  Regarding the metal-insulator
    transition, there is a subtle difference between LR spinless and
    spinfull models that the later seems has no transition for even large
    interactions\cite{fano1999unscreened}.

       Despite large amount of work cited in the above, our understandings of 1D
       LR interacting electrons are still not complete.  
       The widely quoted bosonization formula of near crystalline $4k_F$
       correlation \cite{schulz1993wigner} has been rarely unambiguously
       verified numerically (an exception was in
       \onlinecite{fano1999unscreened}); The discussion of the interplay between
       LR interaction and umklampp scattering  was only made for short
       chains\cite{poilblanc1997insulator}, that need to be checked in large
       systems; 
       In addition, recent advances in cold atom
       experiments, such as polar molecular and trapped
       ions\cite{campbell2015quantum}, have realized quasi-1D LR systems with
       power law interactions $1/r^\alpha$ and exponent $\alpha$ broader than a
       single Coulomb force ($\alpha=1$), which also worth investigating.  Thus
       improved numerical results are still needed.            
       
    In this paper, we study long-range interacting fermions using the \dmrg
    (DMRG) algorithms. 
    The model considered here consists of a chain with $N$ sites, 
    \begin{equation}
        \hat H = \sum\limits_i { - t(\hat c_i^ \dag {{\hat c}_{i{\rm{ + 1}}}} + h.c.) + \sum\limits_{i < j} 
        {{\mathscr{V}(|i-j|)}({{\hat n}_i} - n)({{\hat n}_j} - n)} } 
        \label{eq:ham}
        \end{equation}
    where $\hat c^\dagger_i$ ($\hat c_i$) is creation (annihilation) operator of
    a spinless fermion, $\hat n_i$ is a fermion density operator at site $i$,
    and $n$ is the average density. The interaction potential decays in a power
    law $\V(r)=V/r^\alpha$. The overall amplitude $V$ is
    non-negative representing repulsion force. 
        And the exponent $\alpha$ ranges from $\infty$ to $0$,
        interpolating continuously nearest neighbour interaction to the
        unphysical limit of undecayed interaction. 
        
    We aim to give a systematic account of the interaction range effects and
    determine a phase diagram for the full parameter space.  
    The SR limit
    ($\alpha=\infty$) of Eq.\eqref{eq:ham} is already well understood. We are
    particularly interested in whether the Luttinger liquid theory effects for
    power law interaction when $\alpha<\infty$. According to
    bosonization analysis this should be true if $\sum_r{\V(r)}$ is
    finite\cite{giamarchi2004quantum}, direct numerical evidence is still
    beneficial, though.  We show that when $\alpha$ is large enough, there is indeed a
    Luttinger liquid region and its boundary with insulator phase can be
    determined by the limiting value of the Luttinger parameter. 
    Whereas when $\alpha$ is small, the system clearly deviates from Luttinger
    liquid, and the metal-insulator transition can no longer be determined by
    the Luttinger parameter. In the metallic region the ground
    state has very slowly decayed density correlation, which is qualitatively
    consistent with bosonization predictions. However, the numeric data does not
    fully match with the analytical ones. 
    We also deliver
    systematic analysis of the ground state entanglement, which supports
    conclusions drawn from other quantities.

     
        This work is organized as follows. 
        Sec.\ref{sec:method} introduces the model and the numerical method; 
        Secs.\ref{sec:phase:diagram:1:2} and \ref{sec:phase:diagram:1:3}
        determine the phase diagram for band fillings $n=1/2$ and $n=1/3$,
        respectively; 
        Sec.\ref{sec:discussion} compares bosonization with our numerical
        results.  Conclusions are drawn in Sec.\ref{sec:conclusion}.

\section{methods} \label{sec:method}
    Except for the hopping amplitude $t$ (fixed to unity) and filling factor
    $n$, $V$ and $\alpha$ are two main free parameters that we will investigate. 
    Several limiting cases can be noted in the \Valpha plane: The line of $V=0$
    are free fermions. This is also the case for $\alpha=0$. When
    $\alpha=\infty$, Eq.\eqref{eq:ham} degenerates to the $t\textrm{-}V$ model
    (equivalent to the $XXZ$ model), which is exactly solvable: For
    half filling there is a metal-insulator transition at $V_c=2.0$. 
    Besides, the low energy field theory for the \tvmodel (and other similar models with
    longer, but finite range, interactions) is the Luttinger model plus a cosine term
    representing the umklapp scattering, known as the \SG model. 
    The fixed points for this field theory model are controlled by the Luttinger
    parameter $K$. It is metallic (insulating) if $K$ is smaller
    (larger) than a critical value $K_c$. $K_c$ is determined by band filling,
    which, for spinless fermions, equals $\frac{1}{2}$ and $\frac{2}{9}$ at 
    fillings $n=\frac{1}{2}$ and $n=\frac{1}{3}$, respectively. 
    
    For generic $V$ and $\alpha$, accurate ground state properties can only
    be obtained numerically.  We use the DMRG algorithm \cite{white1992density,
    *white1993density} to achieve this. 
    Traditionally, it was difficult to use DMRG to simulate systems with LR
    interactions.
    This problem has been solved for the type of LR interaction decaying in a
    power law in translation invariant systems, due to a technique invented in
    Ref.\onlinecite{crosswhite2008applying}. The basic idea behind of it is to
    approximating power function with the sum of several exponentials, and
    encode the Hamiltonian compactly in MPOs (see
    \onlinecite{crosswhite2008applying, frowis2010tensor} for details). Both 
    DMRG and this technique are quasi-exact. 
    In our simulation, the Hamiltonian \eqref{eq:ham} is represented with the
    just mentioned technique, using $6$ exponentials, and the ground state wave function
    $|\psi\rangle$ is represented with \mpss
    \cite{dukelsky1998equivalence,*verstraete2004density, *mcculloch2007from,
    *schollwock2011density}. Other algorithm details are regular: If not
    otherwise specified, open boundary
    conditions (OBC) is used for system size $N$ ranging from $20$ to $1600$. Physical
    quantities are calculated by extrapolation in $N$. 
    The quality of the wave function is gauged such that the truncation error
    to be under $10^{-9}$ or the variance 
    $v=\langle \psi| \hat H^2| \psi\rangle-(\langle \psi|\hat H | \psi\rangle)^2$ 
    smaller than $10^{-5}$. This requires a bond dimension
    $D$ of the MPS up to 500.  
    
    As we will need to calculate $K$ in different ways, a variant of
    DMRG---the iDMRG algorithm\cite{mcculloch2008infinite} is also used in this
    paper. 
    This algorithm exploits translation invariance and represent the ground
    state in an infinite \mps (iMPS). 
    Since it works directly in thermodynamic
    limit, no extrapolation in $N$ is needed and boundary effect is reduced to
    minimum.
    Simulations are made and compared with different bond dimensions to
    ensure accuracy, and the values of the correlators are averaged over one unit
    cell to enforce better translation invariance. 
    
    We study two band fillings $n=1/2$ and $1/3$ which correspond to Fermi wave
    vectors $k_F=\pi/2$ and $\pi/3$, respectively.  For each of them we consider
    entire range of $V$ and $\alpha$, however, as we will see, $0\leq\alpha\leq
    3.0$ and moderate $V$ are revealing enough in practice, so the actual
    computations is restricted to this range.   
    Our strategy is to analyze starting from the familiar short-range limit
    ($\alpha=\infty$) and see how phases and quantum critical point (QCP) evolve to
    unknown regimes when $\alpha$ is tuned smaller.
    Because the results for $n=1/2$ and $n=1/3$ have certain qualitative differences,
    they are shown separately in below.


\section{$\lowercase{n}=1/2$ phase diagram} \label{sec:phase:diagram:1:2}
        \begin{figure}   
              \centering
              \scalebox{0.55}[0.55]{\includegraphics{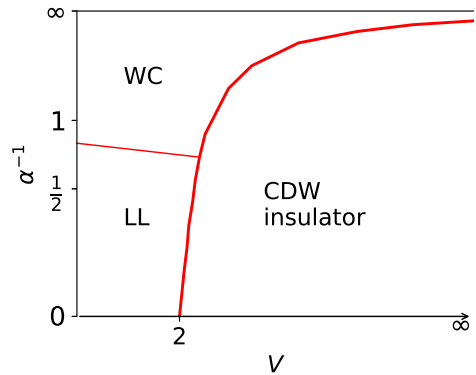}}
              \caption{\label{fig:phase:diagram:1:2} 
              Schematic phase diagram as a function of the interaction amplitude
              $V$ and inverse of exponent $\alpha$ at $n=\frac{1}{2}$.  
              The thick solid line indicates transitions from metallic phases
              to a CDW insulator phase. 
              This line starts from $(V, \alpha)=(2.0, \infty)$ and eventually approaches $\alpha=0$
              when $V\to\infty$.  
              The thin solid line in the metallic regime marks the transition
                from \LL phase (LL) to an \WC phase (WC). 
              In addition, the ground states for $V=0$ or $\alpha=0$ corresponds to 
              free fermions and belong to the \LL phase. 
              }
              \end{figure}

        Fig.\ref{fig:phase:diagram:1:2} shows a schematic phase diagram for
        $n=1/2$. There are three phases: Luttinger liquid, \WC and  CDW phase.
        The former two are  metallic while the later is an insulator. The  
        \LL regime is determined by measuring the Luttinger parameter, as will
        be  explained in subsection \ref{sec:phase:diagram:1:2:K}, where the
        boundary of it with the CDW phase is located accurately, while the
        boundary with the \WC is estimated crudely. The metal-insulator
        transition point is also determined qualitatively in the whole range of
        $\alpha$ by studying the structure factor, charge gap and charge
        stiffness, as will be presented in \ref{sec:phase:diagram:1:2:delta}.
        These resulting phase boundaries are consistent with the entanglement
        properties as will be shown in subsection
        \ref{sec:phase:diagram:1:2:EE}.

\subsection{Luttinger parameter}  \label{sec:phase:diagram:1:2:K}        
    \begin{figure}   
          \centering
          \scalebox{0.55}[0.55]{\includegraphics{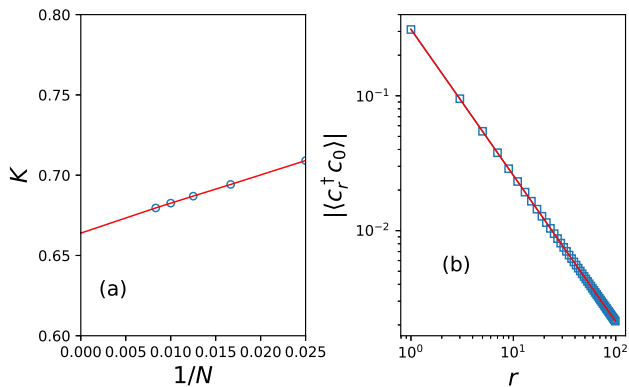}}
          \caption{\label{fig:luttinger:point} 
          Correlation exponent $K$ for $V=1.5$ and $\alpha=2.5$
          extracted through two different methods: 
          (a) By method 1 (see main text): symbols are $K$ evaluated by
          Eq.\eqref{eq:K1} within DMRG versus inverse of system size $N$, for
          $N=40, 60, 80, 100$ and $120$. The line is a fitting to third order
          polynomial in $1/N$. The extrapolated $K$ to thermodynamic limit
          equals $0.6639$. 
          (b) By method 2: symbols are single particle Green's function computed 
          using iDMRG with bound dimension $D=240$. The line is 
          fitting
          of the data to Eq.\eqref{eq:corr:cc} for $32\leq r\leq 60$. Best
          fitting is found at $K=0.6649$.  
          }
          \end{figure}
    
    As mentioned in the introduction, Luttinger liquids are characterized by
    various power law decaying correlations. For example, the connected
    density-density correlation and the single particle Green's function are
    given by the two following formula: 
    \begin{equation}
        \label{eq:corr:nn}
        \langle ({\hat n_r} - \langle {\hat n_r}\rangle )({\hat n_0} - \langle {\hat n_0}\rangle )\rangle  
        = - \frac{K}{{2{\pi ^2}{r^2}}} + 
        {C}\frac{{\cos (2{k_F}r)}}{{{r^{2K}}}} +  \cdots 
    \end{equation}
    \begin{equation}
        \label{eq:corr:cc}
        \langle c_r^\dag {c_0}\rangle \sim {r^{-\frac{1}{2}(K + {K^{ - 1}})}}       
    \end{equation}
    Moreover, the parameter $K$ in the above two lines are the same one. This prominent
    character can be used to detect it. Specifically, one may extract $K$
    numerically from Eqs.\eqref{eq:corr:nn}
    and \eqref{eq:corr:cc}, then it should be in the \LL phase if the two $K$s 
    coincide very well, otherwise it falls outside of this phase.  
    We are meant to use that strategy to demarcate the \LL phase in the
    parameter plane.  This is feasible provided the numerically extracted $K$
    values are enough accurate. Indeed, due to the high accuracy and versatility of the
    DMRG like algorithms, Luttinger parameter can be extracted accurately
    from multiple methods. Two of them which we employ are introduced as
    follows: 
    
    \emph{method 1}. Use DMRG to simulate the ground state and extract $K$ from
    Eq.\eqref{eq:corr:nn}\cite{daul1998ferromagnetic,*clay1999possible,*ejima2005tomonaga}. A direct fitting of the DMRG data with
    Eq.\eqref{eq:corr:nn} will face strong end effects. 
    The usual conduct is to calculate its
    Fourier transform, i.e. the static structure factor 
    \begin{equation}
        \label{eq:structure_factor}
        S(k) = \frac{1}{N}\sum\limits_{i,j} {{e^{ik(i - j)}}(\langle {\hat
        n_i}{\hat n_j}\rangle  - \langle {\hat n_i}\rangle \langle {\hat
        n_j}\rangle )},
    \end{equation}
    then $K$ is given by the slope of $S(k)$ at zero momentum,
     \begin{equation}
        \label{eq:K1}
        K = {\left. {2\pi \frac{{dS(k)}}{{dk}}} \right|_{k = 0}}. 
    \end{equation}
    Numerically, this equation can be approximated by $K = 2\pi [S(2{k_1}) -
    S({k_1})]/{k_1}$, where $k_1=2\pi/N$ is the smallest momentum for size $N$.
    An example of the extrapolated $K$ value for thermodynamic limit in this
    method at $V=1.5$ and $\alpha=2.5$  is shown in
    Fig.\ref{fig:luttinger:point}(a). 
    
    \emph{method 2.} Use iDMRG to calculate the ground state and extract $K$ from
    Eq.\eqref{eq:corr:cc}\cite{karrasch2012luttinger}. Since in the iMPS
    representation the boundary effect is very small, direct fitting of
    $\langle c_r^\dag {c_0}\rangle$ with Eq.\eqref{eq:corr:cc} suffices to get
    accurate $K$.  For small $r$, irrelevant operators may affect the scaling, while for
    large $r$, accuracy loses, so an intermediate fitting region of  $32\leq r \leq60$ has
    been used\cite{kuhner2000one}.  An example of $K$ extracted in this method is shown in
    Fig.\ref{fig:luttinger:point}(b). 

    \begin{figure}   
          \centering
          \scalebox{0.55}[0.55]{\includegraphics{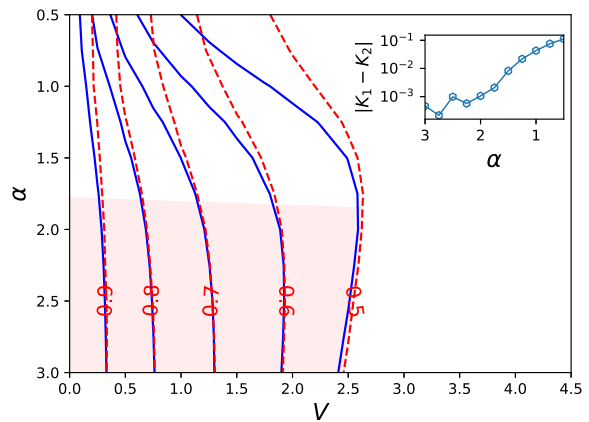}}
          \caption{\label{fig:luttinger:1:2} 
          Contour plot of the correlation exponents $K_1$ (solid) and $K_2$
          (dashed) on the \Valpha plane, with their values restricted to no
          smaller than $K_c=0.5$. Each value of $K_1$ ($K_2$) is 
          calculated using the same way as in Fig.\ref{fig:luttinger:point} (a)
          [(b)]. 
          The shaded region where $K_1$ and $K_2$ are nearly identical marks the
          \LL phase. 
          Inset shows the difference of $K_1$ and $K_2$ vs. $\alpha$ on a
          transect line of $V=1.0$. 
          In this section all results are for $n=1/2$. In the entire paper
          Figs.\ref{fig:luttinger:point}, \ref{fig:luttinger:1:2} and
          \ref{fig:luttinger:1:3} used both DMRG and iMPS, Fig.\ref{fig:K:scale}
          used iMPS and all rests used DMRG solely. 
          }
          \end{figure}
    
    For clarity, we denote the $K$s obtained in the two methods by $K_1$ and
    $K_2$ in order.     
    Contour lines for them in the \Valpha plane are drawn together in
    Fig.\ref{fig:luttinger:1:2}. 
        Good coincidence is found for an extended region in the lower left
        part of the plane. Regarding that $K_1$ and $K_2$ are obtained from different
        algorithms and different quantities, such a coincidence is very remarkable. 
        This means that for large enough $\alpha$ the LR 
        interactions are irrelevant at the Luttinger liquid fixed point and that
        their only effect was a quantitative renormalization of the Luttinger
        liquid parameters. 
        Either $K_1$ or $K_2$ can be taken as the value of
        Luttinger parameter, since they are very close.  Their difference
        $\epsilon\equiv |K_1-K_2|$ can be even taken as an estimate of error,
        which is about $10^{-3}$ for $\alpha\gtrsim2.0$ (see inset of
        Fig.\ref{fig:luttinger:1:2}). 
        Nevertheless, since there is no criterion as to how large an $\epsilon$
        would indicate deviation from Luttinger liquid, it is not able to give
        an accurate $\alpha_c$ for the upper boundary of this region. By direct
        observation, $\alpha_c$ should be between $2.0$ and $1.0$ and depend
        slightly on $V$. A comparison of this result with bosonization arguments
        will be given in section \ref{sec:discussion}.
    
    Taking account of the backward scattering, the above result also implies
    that the \SG model should be still valid for finite $\alpha$ when
    $\alpha>\alpha_c$. Then the lower half of the limiting contour line
    $K_1=0.5$ (or $K_2=0.5$, since they are close and their small discrepancy is
    caused by logarithm corrections which are not incorporated in
    Eqs.\eqref{eq:corr:nn} and \eqref{eq:corr:wc:cc}) determines unambiguously
    how the metal-insulator transition point $V_c$ evolve as $\alpha$ reduces
    from $3.0$ to $\alpha_c$.  For reference, $V_c$ are $2.45\pm0.05$ and
    $2.65\pm0.05$ for $\alpha=3.0$ and $2.0$, respectively. Note that they are
    not far from the critical value of $2.0$ for $\alpha=\infty$.

        %
    
        
    \begin{figure}   
          \centering
          \scalebox{0.55}[0.55]{\includegraphics{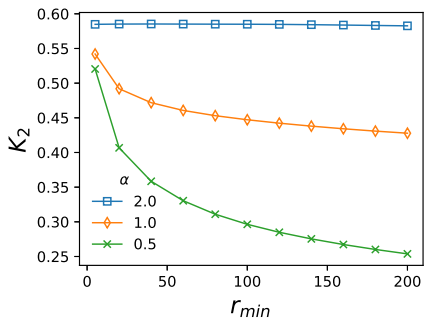}}
          \caption{\label{fig:K:scale} 
          Correlation exponent $K_2$ vs. $r_{min}$ at $V=2.0$ and for several
          values of $\alpha$, where $K_2$ is extracted by fitting iMPS data of
          the single particle Green's function with Eq.\ref{eq:corr:cc} for $r$
          in the range $[r_{min}, r_{min}+30]$ with varying $r_{min}$.  
          }
          \end{figure}
          
    In the small $\alpha$ regime ($0<\alpha<\alpha_c$) large difference between
    the two $K$s indicates the system is in a new fixed point. In this sense,
    this is the true LR regime. 
    Both $K_1$ and $K_2$ quickly reduces as $\alpha$ becomes smaller, indicating
    reinforcement  of the density correlation. 
        Here we emphasize that $K_1$ and $K_2$ are not well defined
        ``Luttinger parameter'' in this regime.  They can still  be
        extracted by curve fitting only because marginal Luttinger liquid character
        is retained.  In fact their values are scale dependent. A demonstration
        of this for $K_2$ is shown in Fig.\ref{fig:K:scale}.  One can see that, at
        $\alpha=2.0$ (in the Luttinger liquid phase), $K_2$ is stable against
        different fitting ranges, meaning Eq.\ref{eq:corr:cc} is very well
        satisfied, while at $\alpha=1.0$ or $0.5$, reduction of $K_2$ with
        increasing length scales indicates deviation from Eq.\ref{eq:corr:cc}.  
    This is another evidence for breakdown of
    Luttinger liquid.     
    Then, considering the backward scattering, the \SG model should be no
    longer justified either. So there is no reason to expect 
    boundary between metallic and insulating phases given by $K_c=0.5$ in  
    this regime. 
    
\subsection{the metal-insulator transition}   \label{sec:phase:diagram:1:2:delta}                

    For large enough $V$ (and $\alpha\neq0$) the potential energy dominates,
    then the fermions are always expected to form an insulating phase. As stated
    in the above, the metal-insulator transition point $V_c$ can be determined
    by Luttinger parameter only in the large $\alpha$ regime.
    In order to see the dependence of the phase boundary on $\alpha$ in its full
    range, in below we {consider some other quantities}. 
    
    \begin{figure}   
          \centering
          \scalebox{0.55}[0.55]{\includegraphics{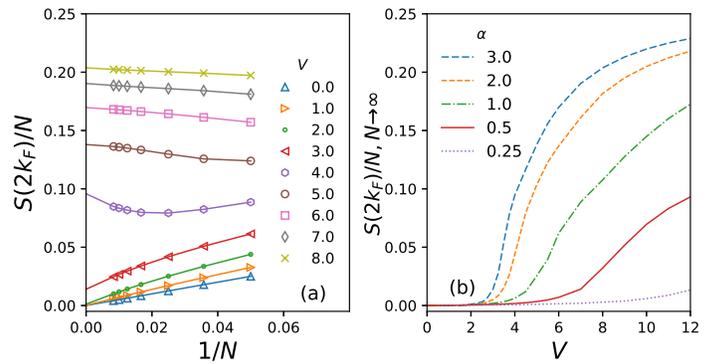}}
          \caption{\label{fig:order_param} 
          (a) Scaled static structure factor at $k=2k_F$ versus inverse system size for
          $\alpha=3.0$ and different $V$.  Lines are fit of the data to third order polynomial in $1/N$.  (b) Extrapolation of $S(2k_F)/N$ to
          infinite $N$ as a function 
          of $V$ for different $\alpha$. 
          }
          \end{figure}
    The metallic phases are translation invariant, while the insulator phase has
    a CDW order with a wave length equal to inverse of the density of fermion
    (corresponding to wave number $k=2k_F$). This $2k_F$ order can be detected
    by a non-vanishing $S(2k_F)/N$ value in thermodynamic limit. 
    The scaling of the structure factor against $1/N$ for $\alpha=3.0$ and
    several $V$ is shown in Fig.\ref{fig:order_param}(a). For $V\leq3.0$, this
    quantity is extrapolated to zero or very small values, while clearly becomes 
    finite for larger $V$.  This behaviour is largely in agreement with
    $V_c=2.45$ obtained by the above study of Luttinger parameter. 
    The extrapolated thermodynamic limit value as a function of $V$ for
    different $\alpha$ is shown in  Fig.\ref{fig:order_param}(b). 
    For each $\alpha$, it changes from zero to finite values at some
    QCP $V_c$, which is not easy to be accurately located. Nevertheless, it 
    is clear that qualitatively the tendency to form CDW order is always
    weakened as $\alpha$ reduces.  
    
   \begin{figure}   
          \centering
          \scalebox{0.55}[0.55]{\includegraphics{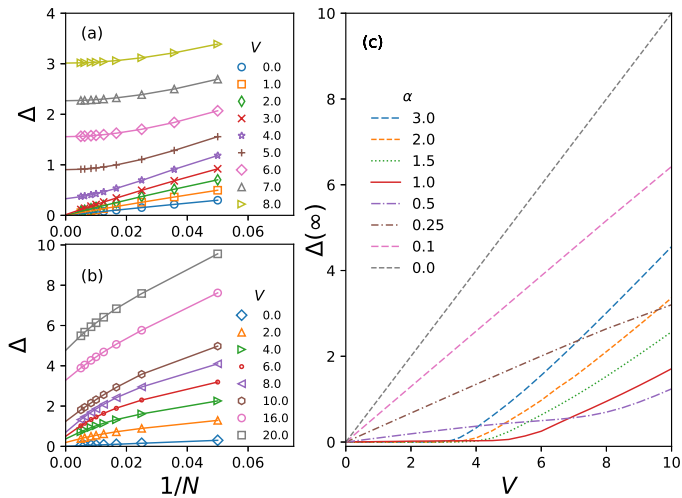}}
          \caption{\label{fig:cg:12} 
          Single particle charge gap $\Delta$ versus inverse system size at
          $\alpha=3.0$ (a)  and $\alpha=0.5$ (b) and for each of them several
          values of $V$. Lines are fit of the data to the function 
          $\Delta(N)= \deltainf + a_1\frac{1}{N} + a_2\frac{1}{N^2} + a_3\frac{1}{N^3}$
          (c) $\deltainf$ extracted from data fitting as a
          function of $V$ for different $\alpha$. 
          }
          \end{figure}
    
    The next quantity considered is the single particle charge
    gap\cite{lieb1968absence}.  In thermodynamic limit it will
    normally be zero in metallic phases, while become finite in insulating phases. Its 
    definition is 
    \begin{equation}
        \label{eq:charge_gap}
        {\Delta}(N) = E({N_f} + 1) + E({N_f} - 1) - 2E({N_f}),
    \end{equation}
    where ${N_f} = nN$ is the number of fermions.  
    Scaling of $\Delta$ with inverse system size at $\alpha=3.0$ and $0.5$ are 
    shown in Fig.\ref{fig:cg:12}(a) and (b), respectively, for different
    interaction strength. Each group of data can be well fitted with third order
    polynomials in $1/N$, leading to certain thermodynamic limit values
    $\deltainf$. Then the extrapolated $\deltainf$ as a function of $V$ are shown for
    different $\alpha$ in Fig.\ref{fig:cg:12}(c), from which, two distinct 
    behaviours  can be observed separated by $\alpha=1$. 
     For each $\alpha>1$, $\deltainf$ changes from zero to finite
    values at certain critical $V_c$, and $V_c$ should get larger as $\alpha$
    reduces; 
    For each $0\leq\alpha\leq 1$, $\deltainf$ is (almost) proportional to $V$ in
    the beginning, and later may still grow linearly in $V$ but with a larger slope
    (At $\alpha=1$ this is marginal and not very obvious, and, at $\alpha=0$,
    $\deltainf = V$ holds exactly for all range of $V$). 

   \begin{figure}   
          \centering
          \scalebox{0.55}[0.55]{\includegraphics{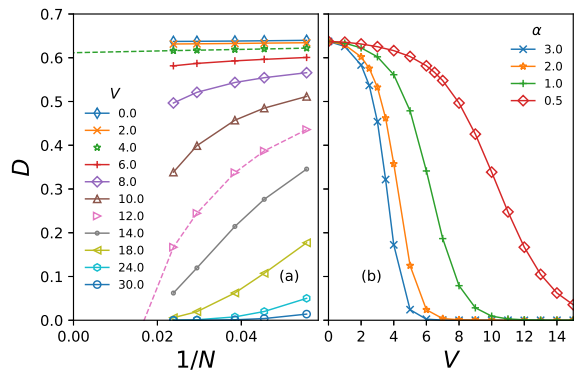}}
          \caption{\label{fig:stiff:12} 
          (a) Charge stiffness $D$  vs. $1/N$ 
          at $\alpha=0.5$ and for different $V$. The system sizes are $N=18, 22, 26,
          34$ and $42$, corresponding to odd numbers of fermions. Solid lines at
          are guide to eye.  Dashed {lines at $V=4.0$ and $12.0$ are fit} to
          third order polynomials in $1/N$. The former fit is good, while the
          later leads to negative $D$ which is clearly incorrect. 
          (b) $D$  as a function of $V$ for different
          $\alpha$, obtained at a fixed finite system size of $N=42$. These data
          are obtained through DMRG with periodic boundary conditions and the bond dimensions used are up to 640. 
          }
          \end{figure}

    The gap for $\alpha>1$ agrees with previous findings about the Luttinger
    parameter and scaled structure factor, while the results for
    $0\leq\alpha\leq 1$ and small $V$ do not: There
    is no $2k_F$ CDW order, yet there is a finite charge gap. This raises a
    question that whether the phase in this regime is a metal
    or insulator?  If we follow the notion of Kohn that metal/insulator should
    be classified by the ground state wave function alone (while not the
    spectrum) \cite{kohn1964theory,resta2002why}, then it should be metallic, 
    and existence of the charge gap should be understood as one peculiar
    feature of true LR interaction potential.
    
    Two facts supporting metallic nature of the $0\leq\alpha\leq1$ and small
    $V$ regime follow. 
    First, consider the exact solvable region of $\alpha=0$. For them the ground
    state wave functions are identical to that of the free fermions of $V=0$,
    and the single particle charge gap can be easily worked
    out to be $\Delta(N)=V+ 4\pi/N + O(1/N^2)$ for a lattice size $N$, which equals $V$ in thermodynamic
    limit. This extreme case provides an exact example of a metal with a finite 
    charge gap. 
    Second, for generic $\alpha$, we measure Kohn's charge stiffness\cite{kohn1964theory}
    \begin{equation}
        D = N{\left. {\frac{{{\partial ^2}E(\phi )}}{{\partial {\phi ^2}}}}
        \right|_{{\phi _m}}}, 
        \label{eq:stiff}
        \end{equation}
    where $\phi$ is a magnetic flux penetrating periodic ring of model
    Eq.\eqref{eq:ham}. $\phi_m$ is the location of minimum of the ground state
    energy, which equals $0$ ($\pi$) for an odd (even) number of fermions.
    Fig.\ref{fig:stiff:12}(a) shows $D$ vs. $1/N$ for fixed
    $\alpha=0.5$ and different $V$, deep in the LR regime, with $N$ no larger than $42$ 
    \footnote{Here multiple factors have severely restricted evaluation of $D$
    accurately for only small $N$: (i) use of periodic boundary conditions, (ii)
    then inapplicability of the MPO technique for approximating LR interactions,
    (iii)
    evaluation of the second order derivative in Eq.\ref{eq:stiff} being very
    numerical delicate.}. 
    When $V$ is small or large, finite size effects are weak and extrapolation
    of $D$ to infinite $N$ should be good. For example, at $V=4.0$, $D$ is
    extrapolated to a finite value, showing a metallic state. 
    Similar conclusions can be reached for other $(V, \alpha)$ pairs when they
    are both small. 
    For intermediate $V$, finite size effects are strong, and extrapolation
    using small system sizes will lead to large error [see the line at $V=12.0$
    of Fig.\ref{fig:stiff:12}(a)]. Then we compare $D$ as a function of $V$ for
    different $\alpha$, with a fixed finite system size of $N=42$, as shown in
    Fig.\ref{fig:stiff:12}(b). The function values are always promoted as
    $\alpha$ reduces, indicating, for the finite system size, stronger metallic
    character for smaller $\alpha$. This result is consistent with the
    behaviour of the extrapolated structure factor shown in
    Fig.\ref{fig:order_param}(b). 
   
    From the above analysis, we conclude that there are metal-insulator
    transitions for all $\alpha>0$ and the transition point $V_c$ shifts to
    larger values as $\alpha$ reduces ($V_c$ should go to infinity  as $\alpha$
    approaches zero). 
    It means that the fermions are less likely to be localized by the
    backscattering by the lattice
    with increasing interaction range. 
    This remarkable fact was uncovered in earlier works of Refs. 
    \onlinecite{poilblanc1997insulator,capponi2000effects} for short chains. We
    confirm this for even larger system sizes and in wider parameter ranges. 
    
    
\subsection{entanglement entropy}  \label{sec:phase:diagram:1:2:EE}
   \begin{figure}   
          \centering
          \scalebox{0.55}[0.55]{\includegraphics{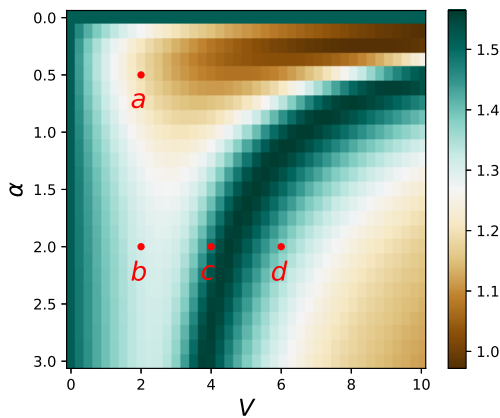}}
          \caption{\label{fig:EE:one_half} 
          Distribution of the half chain entanglement
          entropy $\E_{N/2}$ on the parameter domain
          $(V, \alpha) \in [0, 10]\times[0, 3.0]$ divided into $40\times 24$
          grids for system size $N=100$. Several dots are 
          marked $a, b, c$ and $d$ whose scaling of entanglement are shown in
          Fig.\ref{fig:ee:scaling}.  
          }
          \end{figure}
    In the above, we have determined the approximate ground state phase diagram
    at half filling. 
    It is also interesting to study quantum entanglement
    in the ground states, since it provides alternative characterization of many
    body wave function and may have interesting connection with quantum phase
    transitions\cite{amico2008entanglement}. 
    
    We use the von Neumann
    entanglement entropy as a measure of bipartite entanglement between the subsystems
    residing on the left $L$ sites of the chain and that on the remaining right $N-L$
    sites, which is formulated by 
    \begin{equation}
        \label{eq:ee} 
        \mathcal{E}_{L}=- {\rm{Tr(}}{\rho _L}\log_2 ({\rho _L})), 
        \end{equation}
    and where $\rho_L=\text{Tr}_{ [L+1,N]}(|\psi\rangle\langle\psi|)$ is the reduced
    density matrix of the left subsystem, by tracing out the right one.  
    
    Fig.\ref{fig:EE:one_half} displays an overview of the distribution of half
    chain entanglement $\E_{N/2}$ on the parameter space for $N=100$. 
    Two features can be noted: 
    First, all points on the boundary lines of $V=0$ and $\alpha=0$ take same
    values of entanglement and the values are relatively high, since they are
    identically free fermions.  
    Second, there is an extended peak of entanglement in the middle of the
    figure.  Let us denote by $V_p$ the location of the peak for a given
    $\alpha$. Then the curve $V_p(\alpha)$ has the same trend as $V_c(\alpha)$. This
    supports previous conclusions about the phase diagram, and $V_p$ should have
    intimate connections with $V_c$. Since $V_p$ can be easily obtained, one
    may wonder can it be used to precisely locate the later. 
    In short, the answer is no. 
    For example, at
    $\alpha=3.0$ and $N=100$, $V_p=3.84$ is much larger than $V_c\simeq 2.65$.
    Although $V_p$ relies on $N$ and will scale closer to $V_c$ as system
    size $N$ increases, extrapolation of $V_p$ using very large $N$ can still
    hardly reach $V_c$. (This has been demonstrated for the $XXZ$ model 
    in Ref.\onlinecite{wang2010berezinskii} and persists also for finite
    $\alpha$ which we did not show at here).  
    The reason for the existence of the peak is explained in the next
    paragraphs. 
    We argue that $V_p$ remains a good, strict upper bound  on 
    $V_c$. 
    
    \begin{figure}   
          \centering
          \scalebox{0.55}[0.55]{\includegraphics{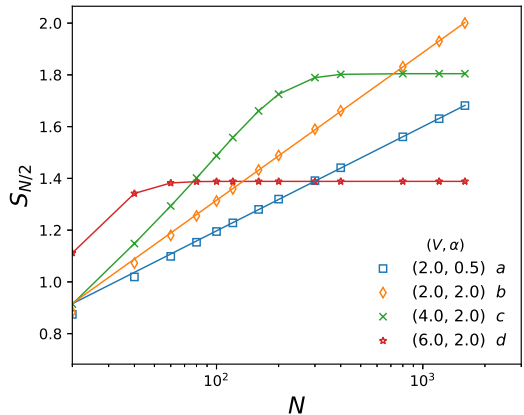}}
          \caption{\label{fig:ee:scaling} 
          Scaling of half chain entanglement $\E_{N/2}$ versus systems size
          $N$ for several $(V, \alpha)$ pairs, with $N$ as large as $1600$. 
          These pairs are labeled $a$, $b$, $c$, and $d$, whose positions are
          also shown in Fig.\ref{fig:EE:one_half}. Lines for $a$ and $b$ are
          fitting to the relation ${\E_{N/2}}\sim\frac{c}{6}{\log _2}(N)$,
          yielding the coefficient $c=0.73$ and $1.03$, respectively.  Lines for $c$ and
          $d$ are guide for eyes. 
          }
          \end{figure}
    Further understanding of entanglement structure of the system can be gained
    by analyzing scaling of $\E_L$ with subsystem size $L$. 
    For gapped and SR systems $\E_L \sim const.$, when $L$ is greater
    than the correlation length. This is the well known area law of
    entanglement for 1D \cite{eisert2010colloquium,hastings2007area}.  Whereas
    for gapless ones, $\E_L$ should (slowly) diverge with increasing $L$. It is
    interesting to check whether this paradigm is altered or not when LR 
    interactions present.
        

    Instead of fixing a $N$ and studying the relation of $\E_L$ with $L$, we
    choose to fix the ratio  $L/N=1/2$ and analyze the scaling of $\E_{N/2}$
    with $N$. The later has the benefit that only one parameter is needed.
    Fig.\ref{fig:ee:scaling} shows the results in a log-linear plot for $4$
    points in the \Valpha plane. They are marked $a$, $b$, $c$ and $d$ in
    Fig.\ref{fig:EE:one_half} and are  representatives of different phases.
    Points $c$ and $d$ are both in the CDW phase, and entanglement saturates for
    large enough $N$ as expected. Point $b$ is in the Luttinger liquid phase,
    and entanglement diverges logarithmically. The interesting case is point $a$
    which is in the WC phase. It has a finite charge gap, but entanglement
    still diverges logarithmically. 
    The two points $b$ and $a$
    have different slope in the logarithmic scalings. One may extract an effective
    central charge for each of them by fitting with the formula
    ${\E_{N/2}}\sim\frac{c}{6}{\log _2}(L)$, which gives $c=1.03$ and $0.73$
    respectively.  The former matches well with the theoretical value of $1$,
    while the later shows again deviation from Luttinger liquid. 
    When entering from the metallic phase (point $b$) to the insulating phase
    (points $c$ and $d$), opening of a small gap leads to faster growth of
    entanglement (comparing $b$ and $c$), while too large gap leads to too soon 
    saturation of entanglement (comparing $c$ and $d$). It is  these two
    competing effects that caused the peak of entanglement near $c$ (for
    $N=100$) in Fig.\ref{fig:EE:one_half}. 
    
    In short, the results of entanglement fully support the
    conclusions ahead. And LR interactions do not lead to very large
    entanglement and more complicated ground states in the model.

\section{\lowercase{$n$}=1/3 phase diagram}  \label{sec:phase:diagram:1:3}
    \begin{figure}   
          \centering
          \scalebox{0.55}[0.55]{\includegraphics{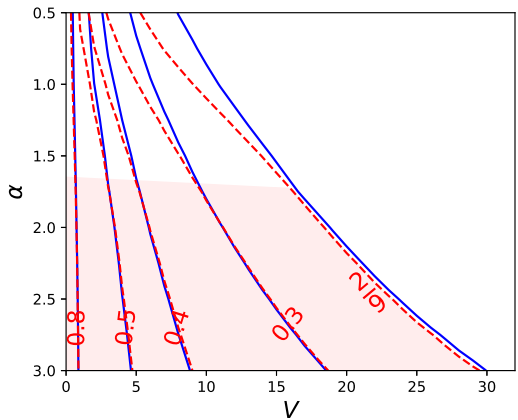}}
          \caption{\label{fig:luttinger:1:3} 
          Contour plot of correlation exponents $K_1$ (solid) and $K_2$ (dashed)
          restricted to $K_{1}, K_2\geq \frac{2}{9}$. They are extracted 
          in the same way as in Fig.\ref{fig:luttinger:point} and
          \ref{fig:luttinger:1:2}. In this section all results are for $n=1/3$.}
          \end{figure}
    
    In this section we study the ground states for filling $n=1/3$. We measure
    some quantities in parallel with $n=1/2$ and compare their results.
    To begin with, the correlation exponents $K_1$
    and $K_2$ are calculated in the same way as before and contour plots of
    them are shown in Fig.\ref{fig:luttinger:1:3}.
    One can see that they agree well in the lower left region of the \Valpha
    plane, demarcating the Luttinger liquid region.
     The $K$ values clearly decrease as $\alpha$ goes down from $3.0$ in this region.
    So, compared with $n=1/2$, LR interactions have more salient effects on
    the renormalized $K$ values. 
    When $\alpha$ is very small, $K_1$ and $K_2$ are obviously different,
    indicating breakdown of Luttinger liquid. It is difficult to determine a critical value
    $\alpha_c$ for when this occurs. Then the boundary between \LL and \WC is
    only roughly estimated to be $1.0\leq\alpha_c<2.0$ and it may depend on $V$.
    The right boundary of \LL phase is determined unambiguously by the critical
    value $K_c=2/9$.  For reference, $V_c$ are $29.8$  
    and $18.2$ for $\alpha=3.0$ and $2.0$, respectively. 

    \begin{figure}   
          \centering
          \scalebox{0.55}[0.55]{\includegraphics{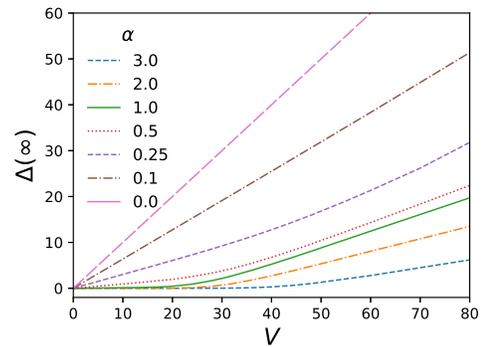}}
          \caption{\label{fig:cg:13} 
          Extrapolated single particle charge gap $\deltainf$ as a function
          of $V$ for different $\alpha$. The values of $\deltainf$ are
          extracted in the same way as that for Fig.\ref{fig:cg:12}. 
          }
          \end{figure}
    
    Next we analyse the metal-insulator transition. 
    For $n=1/2$, the transition has been
    characterized by the occurrence of the CDW order through the scaling of the
    $2k_F$ peak of the structure factor. This is, however, not quite applicable
    for $n=1/3$.  The reason is that there are oscillations (the Friedel
    oscillation) in local density of fermions $\langle \hat n_i \rangle$ in a
    finite size system with OBC. Besides, the oscillation may decay in a power
    law in the metallic phases, which will not damp out for long distances, as a
    result it is difficult to distinguish it with a true CDW order by
    extrapolation in system sizes (we refer to \onlinecite{white2002friedel} for
    detailed discussion of Friedel oscillation for lattice models). 

   \begin{figure}   
          \centering
          \scalebox{0.55}[0.55]{\includegraphics{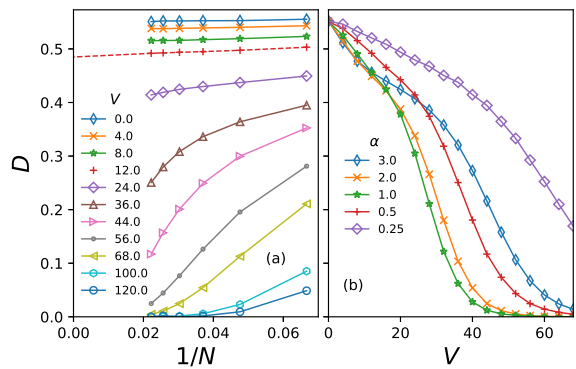}}
          \caption{\label{fig:stiff:13} 
          {(a) Charge stiffness} $D$  vs. $1/N$ 
          at $\alpha=0.5$ and for different interaction strength $V$. 
          The system sizes are {$N=15, 21, 27,
          33, 39$ and $45$}, corresponding to odd numbers of fermions. Solid lines at
          are guide to eye.  Dashed line at $V=12.0$ is fit to a 
          third order polynomial in $1/N$.  
          (b) $D$  as a function of $V$ for different
          $\alpha$, obtained at a fixed finite system {size of $N=45$}. These data
          are obtained through DMRG with periodic boundary conditions and the bond dimensions used are up to 640. 
          }
          \end{figure}
    
    Fig.\ref{fig:cg:13} shows the extrapolated single particle charge gap 
    $\deltainf$ as a function of $V$ for different $\alpha$. Like $n=1/2$,
    two different behaviours appear, which are separated by $\alpha=1$. For each
    $\alpha>1$, $\deltainf$ becomes nonzero at certain $V_c$, which
    reduces for smaller $\alpha$. This shows shrinking of the metallic region as
    interaction range increases. While in the range of $0\leq\alpha\leq1$, it is
    always gapped except at $V=0$; Besides $\deltainf$ is proportional to
    $V$ in the beginning and later increases with $V$ with a larger
    slope for a given $\alpha$ (this is obvious for $\alpha=1.0$ and 0.5,
    but is hidden for smaller $\alpha$ due to restriction of the range of $V$ in the
    figure), which may indicate change from WC to the CDW phase. One
    also notes that $\deltainf=V$ at $\alpha=0$, which is independent of 
    band-filling. 
    
    The gapped region at $0\leq\alpha\leq 1$ and small $V$ is actually metallic.
    This is supported by the study of the charge stiffness, as has been made for
    $n=1/2$. 
    Fig.\ref{fig:stiff:12}(a) shows $D$ vs. $1/N$ for fixed $\alpha=0.5$ and
    different $V$. When $V$ is small or large, finite size effects are weak. An
    extrapolation is performed for $V=12.0$, leading to a finite value, which
    means the state is metallic. Similar conclusions can be reached for other
    parameters of $\alpha$ and $V$ in this region. 
    Fig.\ref{fig:stiff:13}(b) compares dependence  of $D$ on $V$ 
     for different $\alpha$, with a fixed finite system size of $N=45$. 
    For each $\alpha$, $D$ decreases from around the theoretical value
    $\sqrt{3}/{\pi}$ at $V=0$ to zero at some large enough $V$, indicating  a 
    metal-insulator transition. Decrease of $D$ to zero becomes faster
    when $\alpha$ changes from $3.0$ to $1.0$, while becomes slower when
    $\alpha$ continues changing from $1.0$ to $0.25$. 
    This finite size result implies that the metallic region may first shrink and
    then expand as the interaction range increases. 
    
    \begin{figure}   
          \centering  
          \scalebox{0.55}[0.55]{\includegraphics{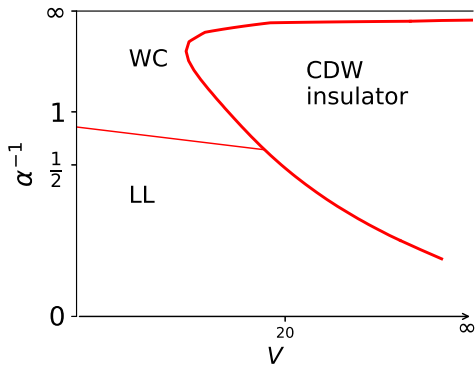}}
          \caption{\label{fig:pd:1:3} 
          Possible phase diagram as a function of $V$ and inverse of $\alpha$
          for $n=1/3$. The thick solid line indicates the boundary between
          metallic phases with the CDW insulator phase. The critical value
          $V_c\to\infty$ as $\alpha$ approaches $0$ or $\infty$. The thin solid
          indicates a transition between the \LL phase (LL) and
          \WC (WC) phase. In addition, entire lines of $V=0$ or $\alpha=0$
          belong to the \LL phase. 
          }
          \end{figure}
    
    \begin{figure}   
          \centering
          \scalebox{0.55}[0.55]{\includegraphics{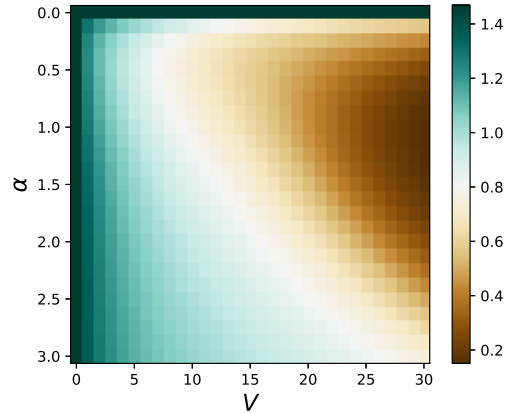}}
          \caption{\label{fig:EE:one_third} 
          Distribution of the half chain entanglement
          entropy $\E_{N/2}$ on the parameter domain
          $(V, \alpha) \in [0, 30]\times[0, 3.0]$ divided into $30 \times 28$
          grids for system size $N=96$. 
          }
          \end{figure}
  
    Combining the above results of the Luttinger parameter, charge gap and
    charge stiffness, one can infer that dependence of the transition point
    $V_c$ on $\alpha$ is not monotonic. $V_c$ should first decrease then, at
    some value of $\alpha$, turns to increase, as $\alpha$ reduces. The location
    of the turning point should be no larger than $1$.  Note also that theoretically there is
    no transition at $\alpha=\infty$ or $\alpha=0$ for for $n=1/3$. Then a
    possible phase diagram is summarized in Fig.\ref{fig:pd:1:3}. 
    It is in essence consistent with the phase diagram for $n=1/2$. The main
    difference is that, in the large $\alpha$ regime, fermions are much harder
    to be localized and LR interactions play a more pronounced role for the
    lower band-filling.

          
    Fig.\ref{fig:EE:one_third} reveals the distribution of half chain
    entanglement on the \Valpha plane.  A convex region in the middle right of
    the figure has lower entanglement, while the rest region on the left has
    higher entanglement. In contrast to $n=1/2$ there is no peak between the two
    regions. The left and right regions should correspond to metallic phases and
    the insulating phase, respectively, so the entanglement distribution correctly
    reflects the phase diagram shown above.

\section{comparison with bosonization formula}  \label{sec:discussion}
    
    The metallic region with small $\alpha$  is of special theoretical interests
    and practical relevance. 
    Although bosonization technique has been extended for 1D fermions with
    Coulomb
    potential\cite{schulz1993wigner,wang2001coulomb,giamarchi2004quantum} and
    general power law
    interactions\cite{iucci2000exact,naon2005conformal,inoue2006conformal}, they
    are not as well established and widely checked as its application in
    short-ranged models.
    So we compare those analytical results with the numerical ones. 
    
    According to arguments from bosonization, for a generic interaction potential, it
    should be essentially short-ranged and falls in the \LL phase if its Fourier
    transform is finite as $q\to0$, otherwise it falls outside that
    phase\cite{giamarchi2004quantum}.
    In particular for the power law potential $\V(r)$, this means a
    critical value at $\alpha_c=1.0$.  Now, comparing with our results, the
    large $\alpha$ region are indeed in the \LL phase, while not for small
    $\alpha$. The difference is that the values of $K_1$ and $K_2$ seems diverge
    before $\alpha$ reaches 1.0, signifying a larger $\alpha_c$ than
    bosonization prediction. But this could also be a phenomenon at a crossover
    scale, and does not exclude the possibility that $K_1$ and $K_2$ may
    coincide for all $\alpha>1.0$ when measured at very large scales. We are not
    conclusive about the accurate location of $\alpha_c$. 
    
    Then we focus on the Coulomb potential $\alpha=1.0$, which is certainly out
    of the Luttinger liquid phase. Main bosonization
    formula are summarized here for comparison. The Fourier transformed
    potential $\V(q\to 0) \to \ln(1/q)$, which leads to a plasmon dispersion, 
    \begin{equation}
        \label{eq:dispersion}
        \omega (q)\sim q{\ln ^{1/2}}(1/q), 
    \end{equation}
    instead of a linear one as the Luttinger liquid. The 
    density-density correlation and single particle Green's function are given respectively
    by the following two lines,
    \begin{equation}
       \label{eq:corr:wc:nn}
       \langle {\hat n_r}{\hat n_0}\rangle \sim \cos (2{k_F}r)\exp (-c_1(\ln r)^{1/2}), 
    \end{equation}
    \begin{equation}
        \label{eq:corr:wc:cc}
        \langle \hat c_r^\dag {\hat c_0}\rangle \sim  \exp ( - c_2{(\ln r)^{3/2}}). 
    \end{equation}
    The former decays slower than any power law and the later faster than any
    power law, which formally correspond to a Luttinger parameter $K\to 0$.
    Eq.\eqref{eq:corr:wc:nn} implies electrons are nearly ordered and can be
    interpreted as a metallic CDW state. 
    In the above, we have shown that the $K$ value reduces as $\alpha$ gets
    smaller (see Figs.\ref{fig:luttinger:1:2} and \ref{fig:K:scale}). Both
    results agree qualitatively on the role of the LR interactions that they
    enhance (diagonal) density correlation while weaken the off-diagonal
    correlation, and that they deviate from standard Luttinger liquids.


    \begin{figure}   
          \centering  
          \scalebox{0.55}[0.55]{\includegraphics{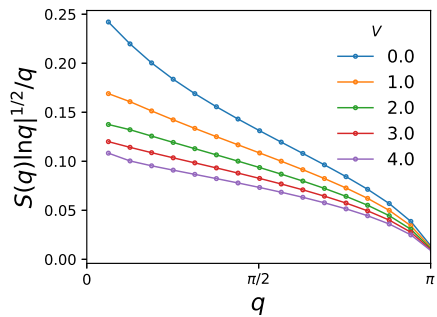}}
          \caption{
          \label{fig:compare} 
          Small momentum behaviour of the structure factor for $\alpha=1.0$ and
          several small values of $V$ at $n=1/2$. 
          }
          \end{figure}

    
    To test the formulas in a quantitative level we follow the approaches
    of Refs.\onlinecite{casula2006ground,shulenburger2008correlation,fano1999unscreened}.  
    They are compared with 
    numerical results for only half filling and small interaction strengths. 
    The dispersion relation can be tested via the
    Feynman formula, which means that the structure factor should behave as $S(q)\sim
    q|\ln q|^{-1/2}$  at small momentum (in contrast to linear relation
    $S(k)\sim qK$ for Luttinger liquid). Then the quantity $S(q)|\ln q|^{1/2}/q$
    would tend to a constant for small $q$. However, this is disapproved by our
    numerical results (see Fig.\ref{fig:compare}). 
    One may then test the density correlation by studying the scaling of the
    peak of structure factor with $N$, which should be well fitted by the
    relation $S(2k_F) \sim N \exp(-c_1\sqrt{\ln N})$, if Eq.\eqref{eq:corr:wc:nn}
    were correct. However, again, we found the fitting are not satisfactory
    (not shown). Given that the density correlation is related to the dispersion
    relation, this is in fact no surprise and they should both hold or fail at
    the same time. 
    
    So we conclude that the bosonization results can only match with our DMRG
    data qualitatively but not quantitatively. 
    
    \begin{figure}   
          \centering  
          \scalebox{0.55}[0.55]{\includegraphics{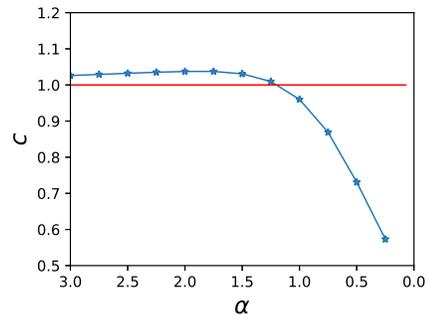}}
          \caption{\label{fig:cc} 
          Effective central charge $c$ as a function of $\alpha$  at 
          $V=2.0$ and $n=1/2$. The $c$ values are extracted in the same way as in
          Fig.\ref{fig:ee:scaling} by fitting the scaling of entanglement with
          the relation ${\E_{N/2}}\sim\frac{c}{6}{\log _2}(N)$. The horizontal
          line of $c=1$ indicates the value of the central charge for the
          Luttinger liquid. 
          }
          \end{figure}
    
    Finally, since some works have been interested in applying conformal field
    theory (CFT) to  the model\cite{naon2005conformal,inoue2006conformal}, for
    reference we show the effective central charge $c$ as a function of $\alpha$
    in Fig.\ref{fig:cc}.  It takes nearly conatant values close to 1 for large
    $\alpha$, indicating preservation of the $U(1)$-invariant CFT of the
    Luttinger liquid, while varies continuously for smaller $\alpha$. This
    behaviour is well consistent with the previous analysis from Luttinger
    parameters.

\section{conclusion}  \label{sec:conclusion}
    
    In summary, we studied the ground states of 1D fermions with LR interaction
    decaying in a power law. 
    We analyzed systematically the effects of interaction ranges on the ground
    state phases and transitions. 
    In the large $\alpha$ regime, the system is essentially short-ranged, where
    the paradigm of \LL theory is still valid and the develop of metal-insulator
    transition point can be determined by the critical value of Luttinger
    parameter; In the small $\alpha$ regime, the \LL theory is breakdown, where
    the ``Luttinger parameter'' is scale dependent. 
    Combining analysis on charge gap, order parameter and entanglement entropy,
    we obtained approximate phase diagram in the whole parameter plane and
    discussed the impact of band filling on it. 
    
    The phase diagrams are in fact not very complicated, in some aspects even
    simpler than certain finite-ranged
    models\cite{mishra2011phase,ejima2005phase}. This relies in the convexity
    of the potential $\V(r)$ we have used, otherwise, other phases, such as the
    bond-order phase, will occur. 
    The main conclusions were reached by comparing the correlation exponents
    extracted accurately in multiple ways within the DMRG/iDMRG techniques, which
    can be also applicable for other models.  These results are relevant to
    ongoing experiments in quasi-1D conductor materials and quantum simulators
    with trapped ions.
    
    The bosonization formula agree with our quasi-exact results qualitatively,
    but not quantitatively. This may either indicates the former can not fully
    capture the low energy behaviour of the current model and elaboration in the
    formulation is needed, or for some other reasons, which needs be clarified
    in future studies on related models.

\section*{acknowledgement}    
     The author thanks prof. An-Min Wang for help with computation
     resources. This work was supported by National Natural Science Foundation
     of China under Grant No. 11375168 and by Research Starting Fund of Xi'an
     Technological University.
    

\section*{References}
    \bibliography{mps-wigner-crystal,mps-wigner-crystal-2}

\begin{thebibliography}{45}%
\makeatletter
\providecommand \@ifxundefined [1]{%
 \@ifx{#1\undefined}
}%
\providecommand \@ifnum [1]{%
 \ifnum #1\expandafter \@firstoftwo
 \else \expandafter \@secondoftwo
 \fi
}%
\providecommand \@ifx [1]{%
 \ifx #1\expandafter \@firstoftwo
 \else \expandafter \@secondoftwo
 \fi
}%
\providecommand \natexlab [1]{#1}%
\providecommand \enquote  [1]{``#1''}%
\providecommand \bibnamefont  [1]{#1}%
\providecommand \bibfnamefont [1]{#1}%
\providecommand \citenamefont [1]{#1}%
\providecommand \href@noop [0]{\@secondoftwo}%
\providecommand \href [0]{\begingroup \@sanitize@url \@href}%
\providecommand \@href[1]{\@@startlink{#1}\@@href}%
\providecommand \@@href[1]{\endgroup#1\@@endlink}%
\providecommand \@sanitize@url [0]{\catcode `\\12\catcode `\$12\catcode
  `\&12\catcode `\#12\catcode `\^12\catcode `\_12\catcode `\%12\relax}%
\providecommand \@@startlink[1]{}%
\providecommand \@@endlink[0]{}%
\providecommand \url  [0]{\begingroup\@sanitize@url \@url }%
\providecommand \@url [1]{\endgroup\@href {#1}{\urlprefix }}%
\providecommand \urlprefix  [0]{URL }%
\providecommand \Eprint [0]{\href }%
\providecommand \doibase [0]{http://dx.doi.org/}%
\providecommand \selectlanguage [0]{\@gobble}%
\providecommand \bibinfo  [0]{\@secondoftwo}%
\providecommand \bibfield  [0]{\@secondoftwo}%
\providecommand \translation [1]{[#1]}%
\providecommand \BibitemOpen [0]{}%
\providecommand \bibitemStop [0]{}%
\providecommand \bibitemNoStop [0]{.\EOS\space}%
\providecommand \EOS [0]{\spacefactor3000\relax}%
\providecommand \BibitemShut  [1]{\csname bibitem#1\endcsname}%
\let\auto@bib@innerbib\@empty
\bibitem [{\citenamefont {Haldane}(1981)}]{haldane1981luttinger}%
  \BibitemOpen
  \bibfield  {author} {\bibinfo {author} {\bibfnamefont {F.~D.~M.}\
  \bibnamefont {Haldane}},\ }\href {\doibase 10.1088/0022-3719/14/19/010}
  {\bibfield  {journal} {\bibinfo  {journal} {J. Phys. C: Solid State Phys.}\
  }\textbf {\bibinfo {volume} {14}},\ \bibinfo {pages} {2585} (\bibinfo {year}
  {1981})}\BibitemShut {NoStop}%
\bibitem [{\citenamefont {Voit}(1995)}]{voit1995one}%
  \BibitemOpen
  \bibfield  {author} {\bibinfo {author} {\bibfnamefont {J.}~\bibnamefont
  {Voit}},\ }\href {\doibase 10.1088/0034-4885/58/9/002} {\bibfield  {journal}
  {\bibinfo  {journal} {Rep. Prog. Phys.}\ }\textbf {\bibinfo {volume} {58}},\
  \bibinfo {pages} {977} (\bibinfo {year} {1995})}\BibitemShut {NoStop}%
\bibitem [{\citenamefont {Giamarchi}(2004)}]{giamarchi2004quantum}%
  \BibitemOpen
  \bibfield  {author} {\bibinfo {author} {\bibfnamefont {T.}~\bibnamefont
  {Giamarchi}},\ }\href@noop {} {\emph {\bibinfo {title} {Quantum physics in
  one dimension}}},\ Vol.\ \bibinfo {volume} {121}\ (\bibinfo  {publisher}
  {Oxford university press},\ \bibinfo {year} {2004})\BibitemShut {NoStop}%
\bibitem [{\citenamefont {Goni}\ \emph {et~al.}(1991)\citenamefont {Goni},
  \citenamefont {Pinczuk}, \citenamefont {Weiner}, \citenamefont {Calleja},
  \citenamefont {Dennis}, \citenamefont {Pfeiffer},\ and\ \citenamefont
  {West}}]{goni1991one}%
  \BibitemOpen
  \bibfield  {author} {\bibinfo {author} {\bibfnamefont {A.~R.}\ \bibnamefont
  {Goni}}, \bibinfo {author} {\bibfnamefont {A.}~\bibnamefont {Pinczuk}},
  \bibinfo {author} {\bibfnamefont {J.~S.}\ \bibnamefont {Weiner}}, \bibinfo
  {author} {\bibfnamefont {J.~M.}\ \bibnamefont {Calleja}}, \bibinfo {author}
  {\bibfnamefont {B.~S.}\ \bibnamefont {Dennis}}, \bibinfo {author}
  {\bibfnamefont {L.~N.}\ \bibnamefont {Pfeiffer}}, \ and\ \bibinfo {author}
  {\bibfnamefont {K.~W.}\ \bibnamefont {West}},\ }\href {\doibase
  10.1103/PhysRevLett.67.3298} {\bibfield  {journal} {\bibinfo  {journal}
  {Phys. Rev. Lett.}\ }\textbf {\bibinfo {volume} {67}},\ \bibinfo {pages}
  {3298} (\bibinfo {year} {1991})}\BibitemShut {NoStop}%
\bibitem [{\citenamefont {Steinberg}\ \emph {et~al.}(2006)\citenamefont
  {Steinberg}, \citenamefont {Auslaender}, \citenamefont {Yacoby},
  \citenamefont {Qian}, \citenamefont {Fiete}, \citenamefont {Tserkovnyak},
  \citenamefont {Halperin}, \citenamefont {Baldwin}, \citenamefont {Pfeiffer},\
  and\ \citenamefont {West}}]{steinberg2006localization}%
  \BibitemOpen
  \bibfield  {author} {\bibinfo {author} {\bibfnamefont {H.}~\bibnamefont
  {Steinberg}}, \bibinfo {author} {\bibfnamefont {O.~M.}\ \bibnamefont
  {Auslaender}}, \bibinfo {author} {\bibfnamefont {A.}~\bibnamefont {Yacoby}},
  \bibinfo {author} {\bibfnamefont {J.}~\bibnamefont {Qian}}, \bibinfo {author}
  {\bibfnamefont {G.~A.}\ \bibnamefont {Fiete}}, \bibinfo {author}
  {\bibfnamefont {Y.}~\bibnamefont {Tserkovnyak}}, \bibinfo {author}
  {\bibfnamefont {B.~I.}\ \bibnamefont {Halperin}}, \bibinfo {author}
  {\bibfnamefont {K.~W.}\ \bibnamefont {Baldwin}}, \bibinfo {author}
  {\bibfnamefont {L.~N.}\ \bibnamefont {Pfeiffer}}, \ and\ \bibinfo {author}
  {\bibfnamefont {K.~W.}\ \bibnamefont {West}},\ }\href {\doibase
  10.1103/PhysRevB.73.113307} {\bibfield  {journal} {\bibinfo  {journal} {Phys.
  Rev. B}\ }\textbf {\bibinfo {volume} {73}},\ \bibinfo {pages} {113307}
  (\bibinfo {year} {2006})}\BibitemShut {NoStop}%
\bibitem [{\citenamefont {Deshpande}\ and\ \citenamefont
  {Bockrath}(2008)}]{deshpande2008one}%
  \BibitemOpen
  \bibfield  {author} {\bibinfo {author} {\bibfnamefont {V.~V.}\ \bibnamefont
  {Deshpande}}\ and\ \bibinfo {author} {\bibfnamefont {M.}~\bibnamefont
  {Bockrath}},\ }\href {\doibase 10.1038/nphys895} {\bibfield  {journal}
  {\bibinfo  {journal} {Nature Physics}\ }\textbf {\bibinfo {volume} {4}},\
  \bibinfo {pages} {314} (\bibinfo {year} {2008})}\BibitemShut {NoStop}%
\bibitem [{\citenamefont {Schulz}(1993)}]{schulz1993wigner}%
  \BibitemOpen
  \bibfield  {author} {\bibinfo {author} {\bibfnamefont {H.~J.}\ \bibnamefont
  {Schulz}},\ }\href {\doibase 10.1103/PhysRevLett.71.1864} {\bibfield
  {journal} {\bibinfo  {journal} {Phys. Rev. Lett.}\ }\textbf {\bibinfo
  {volume} {71}},\ \bibinfo {pages} {1864} (\bibinfo {year}
  {1993})}\BibitemShut {NoStop}%
\bibitem [{\citenamefont {Casula}\ \emph {et~al.}(2006)\citenamefont {Casula},
  \citenamefont {Sorella},\ and\ \citenamefont {Senatore}}]{casula2006ground}%
  \BibitemOpen
  \bibfield  {author} {\bibinfo {author} {\bibfnamefont {M.}~\bibnamefont
  {Casula}}, \bibinfo {author} {\bibfnamefont {S.}~\bibnamefont {Sorella}}, \
  and\ \bibinfo {author} {\bibfnamefont {G.}~\bibnamefont {Senatore}},\ }\href
  {\doibase 10.1103/PhysRevB.74.245427} {\bibfield  {journal} {\bibinfo
  {journal} {Phys. Rev. B}\ }\textbf {\bibinfo {volume} {74}},\ \bibinfo
  {pages} {245427} (\bibinfo {year} {2006})}\BibitemShut {NoStop}%
\bibitem [{\citenamefont {Shulenburger}\ \emph {et~al.}(2008)\citenamefont
  {Shulenburger}, \citenamefont {Casula}, \citenamefont {Senatore},\ and\
  \citenamefont {Martin}}]{shulenburger2008correlation}%
  \BibitemOpen
  \bibfield  {author} {\bibinfo {author} {\bibfnamefont {L.}~\bibnamefont
  {Shulenburger}}, \bibinfo {author} {\bibfnamefont {M.}~\bibnamefont
  {Casula}}, \bibinfo {author} {\bibfnamefont {G.}~\bibnamefont {Senatore}}, \
  and\ \bibinfo {author} {\bibfnamefont {R.~M.}\ \bibnamefont {Martin}},\
  }\href {\doibase 10.1103/PhysRevB.78.165303} {\bibfield  {journal} {\bibinfo
  {journal} {Phys. Rev. B}\ }\textbf {\bibinfo {volume} {78}},\ \bibinfo
  {pages} {165303} (\bibinfo {year} {2008})}\BibitemShut {NoStop}%
\bibitem [{\citenamefont {Astrakharchik}\ and\ \citenamefont
  {Girardeau}(2011)}]{astrakharchik2011exact}%
  \BibitemOpen
  \bibfield  {author} {\bibinfo {author} {\bibfnamefont {G.~E.}\ \bibnamefont
  {Astrakharchik}}\ and\ \bibinfo {author} {\bibfnamefont {M.~D.}\ \bibnamefont
  {Girardeau}},\ }\href {\doibase 10.1103/PhysRevB.83.153303} {\bibfield
  {journal} {\bibinfo  {journal} {Physical Review B}\ }\textbf {\bibinfo
  {volume} {83}},\ \bibinfo {pages} {153303} (\bibinfo {year} {2011})},\
  \bibinfo {note} {arXiv: 1101.0103}\BibitemShut {NoStop}%
\bibitem [{\citenamefont {Ferr{\'e}}\ \emph {et~al.}(2015)\citenamefont
  {Ferr{\'e}}, \citenamefont {Astrakharchik},\ and\ \citenamefont
  {Boronat}}]{ferre2015phase}%
  \BibitemOpen
  \bibfield  {author} {\bibinfo {author} {\bibfnamefont {G.}~\bibnamefont
  {Ferr{\'e}}}, \bibinfo {author} {\bibfnamefont {G.~E.}\ \bibnamefont
  {Astrakharchik}}, \ and\ \bibinfo {author} {\bibfnamefont {J.}~\bibnamefont
  {Boronat}},\ }\href {\doibase 10.1103/PhysRevB.92.245305} {\bibfield
  {journal} {\bibinfo  {journal} {Phys. Rev. B}\ }\textbf {\bibinfo {volume}
  {92}},\ \bibinfo {pages} {245305} (\bibinfo {year} {2015})}\BibitemShut
  {NoStop}%
\bibitem [{\citenamefont {Fano}\ \emph {et~al.}(1999)\citenamefont {Fano},
  \citenamefont {Ortolani}, \citenamefont {Parola},\ and\ \citenamefont
  {Ziosi}}]{fano1999unscreened}%
  \BibitemOpen
  \bibfield  {author} {\bibinfo {author} {\bibfnamefont {G.}~\bibnamefont
  {Fano}}, \bibinfo {author} {\bibfnamefont {F.}~\bibnamefont {Ortolani}},
  \bibinfo {author} {\bibfnamefont {A.}~\bibnamefont {Parola}}, \ and\ \bibinfo
  {author} {\bibfnamefont {L.}~\bibnamefont {Ziosi}},\ }\href {\doibase
  10.1103/PhysRevB.60.15654} {\bibfield  {journal} {\bibinfo  {journal} {Phys.
  Rev. B}\ }\textbf {\bibinfo {volume} {60}},\ \bibinfo {pages} {15654}
  (\bibinfo {year} {1999})}\BibitemShut {NoStop}%
\bibitem [{\citenamefont {Hohenadler}\ \emph {et~al.}(2012)\citenamefont
  {Hohenadler}, \citenamefont {Wessel}, \citenamefont {Daghofer},\ and\
  \citenamefont {Assaad}}]{hohenadler2012interaction}%
  \BibitemOpen
  \bibfield  {author} {\bibinfo {author} {\bibfnamefont {M.}~\bibnamefont
  {Hohenadler}}, \bibinfo {author} {\bibfnamefont {S.}~\bibnamefont {Wessel}},
  \bibinfo {author} {\bibfnamefont {M.}~\bibnamefont {Daghofer}}, \ and\
  \bibinfo {author} {\bibfnamefont {F.~F.}\ \bibnamefont {Assaad}},\ }\href
  {\doibase 10.1103/PhysRevB.85.195115} {\bibfield  {journal} {\bibinfo
  {journal} {Phys. Rev. B}\ }\textbf {\bibinfo {volume} {85}},\ \bibinfo
  {pages} {195115} (\bibinfo {year} {2012})}\BibitemShut {NoStop}%
\bibitem [{\citenamefont {Poilblanc}\ \emph {et~al.}(1997)\citenamefont
  {Poilblanc}, \citenamefont {Yunoki}, \citenamefont {Maekawa},\ and\
  \citenamefont {Dagotto}}]{poilblanc1997insulator}%
  \BibitemOpen
  \bibfield  {author} {\bibinfo {author} {\bibfnamefont {D.}~\bibnamefont
  {Poilblanc}}, \bibinfo {author} {\bibfnamefont {S.}~\bibnamefont {Yunoki}},
  \bibinfo {author} {\bibfnamefont {S.}~\bibnamefont {Maekawa}}, \ and\
  \bibinfo {author} {\bibfnamefont {E.}~\bibnamefont {Dagotto}},\ }\href
  {\doibase 10.1103/PhysRevB.56.R1645} {\bibfield  {journal} {\bibinfo
  {journal} {Phys. Rev. B}\ }\textbf {\bibinfo {volume} {56}},\ \bibinfo
  {pages} {R1645} (\bibinfo {year} {1997})}\BibitemShut {NoStop}%
\bibitem [{\citenamefont {Capponi}\ \emph {et~al.}(2000)\citenamefont
  {Capponi}, \citenamefont {Poilblanc},\ and\ \citenamefont
  {Giamarchi}}]{capponi2000effects}%
  \BibitemOpen
  \bibfield  {author} {\bibinfo {author} {\bibfnamefont {S.}~\bibnamefont
  {Capponi}}, \bibinfo {author} {\bibfnamefont {D.}~\bibnamefont {Poilblanc}},
  \ and\ \bibinfo {author} {\bibfnamefont {T.}~\bibnamefont {Giamarchi}},\
  }\href {\doibase 10.1103/PhysRevB.61.13410} {\bibfield  {journal} {\bibinfo
  {journal} {Phys. Rev. B}\ }\textbf {\bibinfo {volume} {61}},\ \bibinfo
  {pages} {13410} (\bibinfo {year} {2000})}\BibitemShut {NoStop}%
\bibitem [{\citenamefont {Campbell}\ \emph {et~al.}(2015)\citenamefont
  {Campbell}, \citenamefont {Monroe}, \citenamefont {Edwards}, \citenamefont
  {Islam}, \citenamefont {Kafri}, \citenamefont {Korenblit}, \citenamefont
  {Lee}, \citenamefont {Richerme}, \citenamefont {Senko},\ and\ \citenamefont
  {Smith}}]{campbell2015quantum}%
  \BibitemOpen
  \bibfield  {author} {\bibinfo {author} {\bibfnamefont {W.~C.}\ \bibnamefont
  {Campbell}}, \bibinfo {author} {\bibfnamefont {C.}~\bibnamefont {Monroe}},
  \bibinfo {author} {\bibfnamefont {E.~E.}\ \bibnamefont {Edwards}}, \bibinfo
  {author} {\bibfnamefont {R.}~\bibnamefont {Islam}}, \bibinfo {author}
  {\bibfnamefont {D.}~\bibnamefont {Kafri}}, \bibinfo {author} {\bibfnamefont
  {S.}~\bibnamefont {Korenblit}}, \bibinfo {author} {\bibfnamefont
  {A.}~\bibnamefont {Lee}}, \bibinfo {author} {\bibfnamefont {P.}~\bibnamefont
  {Richerme}}, \bibinfo {author} {\bibfnamefont {C.}~\bibnamefont {Senko}}, \
  and\ \bibinfo {author} {\bibfnamefont {J.}~\bibnamefont {Smith}},\ }\href
  {http://escholarship.org/uc/item/2kp9x84m} {\bibfield  {journal} {\bibinfo
  {journal} {Ion Traps for Tomorrow’s Applications}\ }\textbf {\bibinfo
  {volume} {189}},\ \bibinfo {pages} {169} (\bibinfo {year}
  {2015})}\BibitemShut {NoStop}%
\bibitem [{\citenamefont {White}(1992)}]{white1992density}%
  \BibitemOpen
  \bibfield  {author} {\bibinfo {author} {\bibfnamefont {S.~R.}\ \bibnamefont
  {White}},\ }\href {\doibase 10.1103/PhysRevLett.69.2863} {\bibfield
  {journal} {\bibinfo  {journal} {Phys. Rev. Lett.}\ }\textbf {\bibinfo
  {volume} {69}},\ \bibinfo {pages} {2863} (\bibinfo {year}
  {1992})}\BibitemShut {NoStop}%
\bibitem [{\citenamefont {White}(1993)}]{white1993density}%
  \BibitemOpen
  \bibfield  {author} {\bibinfo {author} {\bibfnamefont {S.~R.}\ \bibnamefont
  {White}},\ }\href {\doibase 10.1103/PhysRevB.48.10345} {\bibfield  {journal}
  {\bibinfo  {journal} {Phys. Rev. B}\ }\textbf {\bibinfo {volume} {48}},\
  \bibinfo {pages} {10345} (\bibinfo {year} {1993})}\BibitemShut {NoStop}%
\bibitem [{\citenamefont {Crosswhite}\ \emph {et~al.}(2008)\citenamefont
  {Crosswhite}, \citenamefont {Doherty},\ and\ \citenamefont
  {Vidal}}]{crosswhite2008applying}%
  \BibitemOpen
  \bibfield  {author} {\bibinfo {author} {\bibfnamefont {G.~M.}\ \bibnamefont
  {Crosswhite}}, \bibinfo {author} {\bibfnamefont {A.~C.}\ \bibnamefont
  {Doherty}}, \ and\ \bibinfo {author} {\bibfnamefont {G.}~\bibnamefont
  {Vidal}},\ }\href {\doibase 10.1103/PhysRevB.78.035116} {\bibfield  {journal}
  {\bibinfo  {journal} {Phys. Rev. B}\ }\textbf {\bibinfo {volume} {78}},\
  \bibinfo {pages} {035116} (\bibinfo {year} {2008})}\BibitemShut {NoStop}%
\bibitem [{\citenamefont {Fr\"{o}wis}\ \emph {et~al.}(2010)\citenamefont
  {Fr\"{o}wis}, \citenamefont {Nebendahl},\ and\ \citenamefont
  {D\"{u}r}}]{frowis2010tensor}%
  \BibitemOpen
  \bibfield  {author} {\bibinfo {author} {\bibfnamefont {F.}~\bibnamefont
  {Fr\"{o}wis}}, \bibinfo {author} {\bibfnamefont {V.}~\bibnamefont
  {Nebendahl}}, \ and\ \bibinfo {author} {\bibfnamefont {W.}~\bibnamefont
  {D\"{u}r}},\ }\href {\doibase 10.1103/PhysRevA.81.062337} {\bibfield
  {journal} {\bibinfo  {journal} {Phys. Rev. A}\ }\textbf {\bibinfo {volume}
  {81}},\ \bibinfo {pages} {062337} (\bibinfo {year} {2010})}\BibitemShut
  {NoStop}%
\bibitem [{\citenamefont {Dukelsky}\ \emph {et~al.}(1998)\citenamefont
  {Dukelsky}, \citenamefont {Mart\'{i}n-Delgado}, \citenamefont {Nishino},\
  and\ \citenamefont {Sierra}}]{dukelsky1998equivalence}%
  \BibitemOpen
  \bibfield  {author} {\bibinfo {author} {\bibfnamefont {J.}~\bibnamefont
  {Dukelsky}}, \bibinfo {author} {\bibfnamefont {M.~A.}\ \bibnamefont
  {Mart\'{i}n-Delgado}}, \bibinfo {author} {\bibfnamefont {T.}~\bibnamefont
  {Nishino}}, \ and\ \bibinfo {author} {\bibfnamefont {G.}~\bibnamefont
  {Sierra}},\ }\href {\doibase 10.1209/epl/i1998-00381-x} {\bibfield  {journal}
  {\bibinfo  {journal} {Europhys. Lett.}\ }\textbf {\bibinfo {volume} {43}},\
  \bibinfo {pages} {457} (\bibinfo {year} {1998})}\BibitemShut {NoStop}%
\bibitem [{\citenamefont {Verstraete}\ \emph {et~al.}(2004)\citenamefont
  {Verstraete}, \citenamefont {Porras},\ and\ \citenamefont
  {Cirac}}]{verstraete2004density}%
  \BibitemOpen
  \bibfield  {author} {\bibinfo {author} {\bibfnamefont {F.}~\bibnamefont
  {Verstraete}}, \bibinfo {author} {\bibfnamefont {D.}~\bibnamefont {Porras}},
  \ and\ \bibinfo {author} {\bibfnamefont {J.~I.}\ \bibnamefont {Cirac}},\
  }\href {\doibase 10.1103/PhysRevLett.93.227205} {\bibfield  {journal}
  {\bibinfo  {journal} {Phys. Rev. Lett.}\ }\textbf {\bibinfo {volume} {93}},\
  \bibinfo {pages} {227205} (\bibinfo {year} {2004})}\BibitemShut {NoStop}%
\bibitem [{\citenamefont {{McCulloch}}(2007)}]{mcculloch2007from}%
  \BibitemOpen
  \bibfield  {author} {\bibinfo {author} {\bibfnamefont {I.~P.}\ \bibnamefont
  {{McCulloch}}},\ }\href {\doibase 10.1088/1742-5468/2007/10/P10014}
  {\bibfield  {journal} {\bibinfo  {journal} {J. Stat. Mech. Theor. Exp.}\
  }\textbf {\bibinfo {volume} {2007}},\ \bibinfo {pages} {P10014} (\bibinfo
  {year} {2007})}\BibitemShut {NoStop}%
\bibitem [{\citenamefont {Schollw\"{o}ck}(2011)}]{schollwock2011density}%
  \BibitemOpen
  \bibfield  {author} {\bibinfo {author} {\bibfnamefont {U.}~\bibnamefont
  {Schollw\"{o}ck}},\ }\href {\doibase
  http://dx.doi.org/10.1016/j.aop.2010.09.012} {\bibfield  {journal} {\bibinfo
  {journal} {Ann. Phys.}\ }\textbf {\bibinfo {volume} {326}},\ \bibinfo {pages}
  {96} (\bibinfo {year} {2011})}\BibitemShut {NoStop}%
\bibitem [{\citenamefont {McCulloch}(2008)}]{mcculloch2008infinite}%
  \BibitemOpen
  \bibfield  {author} {\bibinfo {author} {\bibfnamefont {I.~P.}\ \bibnamefont
  {McCulloch}},\ }\href {http://arxiv.org/abs/0804.2509} {\bibfield  {journal}
  {\bibinfo  {journal} {{arXiv}:0804.2509 [cond-mat]}\ } (\bibinfo {year}
  {2008})}\BibitemShut {NoStop}%
\bibitem [{\citenamefont {Daul}\ and\ \citenamefont
  {Noack}(1998)}]{daul1998ferromagnetic}%
  \BibitemOpen
  \bibfield  {author} {\bibinfo {author} {\bibfnamefont {S.}~\bibnamefont
  {Daul}}\ and\ \bibinfo {author} {\bibfnamefont {R.~M.}\ \bibnamefont
  {Noack}},\ }\href {\doibase 10.1103/PhysRevB.58.2635} {\bibfield  {journal}
  {\bibinfo  {journal} {Phys. Rev. B}\ }\textbf {\bibinfo {volume} {58}},\
  \bibinfo {pages} {2635} (\bibinfo {year} {1998})}\BibitemShut {NoStop}%
\bibitem [{\citenamefont {Clay}\ \emph {et~al.}(1999)\citenamefont {Clay},
  \citenamefont {Sandvik},\ and\ \citenamefont {Campbell}}]{clay1999possible}%
  \BibitemOpen
  \bibfield  {author} {\bibinfo {author} {\bibfnamefont {R.~T.}\ \bibnamefont
  {Clay}}, \bibinfo {author} {\bibfnamefont {A.~W.}\ \bibnamefont {Sandvik}}, \
  and\ \bibinfo {author} {\bibfnamefont {D.~K.}\ \bibnamefont {Campbell}},\
  }\href {\doibase 10.1103/PhysRevB.59.4665} {\bibfield  {journal} {\bibinfo
  {journal} {Phys. Rev. B}\ }\textbf {\bibinfo {volume} {59}},\ \bibinfo
  {pages} {4665} (\bibinfo {year} {1999})}\BibitemShut {NoStop}%
\bibitem [{\citenamefont {Ejima}\ \emph
  {et~al.}(2005{\natexlab{a}})\citenamefont {Ejima}, \citenamefont {Gebhard},\
  and\ \citenamefont {Nishimoto}}]{ejima2005tomonaga}%
  \BibitemOpen
  \bibfield  {author} {\bibinfo {author} {\bibfnamefont {S.}~\bibnamefont
  {Ejima}}, \bibinfo {author} {\bibfnamefont {F.}~\bibnamefont {Gebhard}}, \
  and\ \bibinfo {author} {\bibfnamefont {S.}~\bibnamefont {Nishimoto}},\ }\href
  {\doibase 10.1209/epl/i2005-10020-8} {\bibfield  {journal} {\bibinfo
  {journal} {Europhysics Letters (EPL)}\ }\textbf {\bibinfo {volume} {70}},\
  \bibinfo {pages} {492} (\bibinfo {year} {2005}{\natexlab{a}})}\BibitemShut
  {NoStop}%
\bibitem [{\citenamefont {Karrasch}\ and\ \citenamefont
  {Moore}(2012)}]{karrasch2012luttinger}%
  \BibitemOpen
  \bibfield  {author} {\bibinfo {author} {\bibfnamefont {C.}~\bibnamefont
  {Karrasch}}\ and\ \bibinfo {author} {\bibfnamefont {J.~E.}\ \bibnamefont
  {Moore}},\ }\href {\doibase 10.1103/PhysRevB.86.155156} {\bibfield  {journal}
  {\bibinfo  {journal} {Phys. Rev. B}\ }\textbf {\bibinfo {volume} {86}},\
  \bibinfo {pages} {155156} (\bibinfo {year} {2012})}\BibitemShut {NoStop}%
\bibitem [{\citenamefont {K{\"u}hner}\ \emph {et~al.}(2000)\citenamefont
  {K{\"u}hner}, \citenamefont {White},\ and\ \citenamefont
  {Monien}}]{kuhner2000one}%
  \BibitemOpen
  \bibfield  {author} {\bibinfo {author} {\bibfnamefont {T.~D.}\ \bibnamefont
  {K{\"u}hner}}, \bibinfo {author} {\bibfnamefont {S.~R.}\ \bibnamefont
  {White}}, \ and\ \bibinfo {author} {\bibfnamefont {H.}~\bibnamefont
  {Monien}},\ }\href {\doibase 10.1103/PhysRevB.61.12474} {\bibfield  {journal}
  {\bibinfo  {journal} {Phys. Rev. B}\ }\textbf {\bibinfo {volume} {61}},\
  \bibinfo {pages} {12474} (\bibinfo {year} {2000})}\BibitemShut {NoStop}%
\bibitem [{\citenamefont {Lieb}\ and\ \citenamefont
  {Wu}(1968)}]{lieb1968absence}%
  \BibitemOpen
  \bibfield  {author} {\bibinfo {author} {\bibfnamefont {E.~H.}\ \bibnamefont
  {Lieb}}\ and\ \bibinfo {author} {\bibfnamefont {F.~Y.}\ \bibnamefont {Wu}},\
  }\href {https://journals.aps.org/prl/abstract/10.1103/PhysRevLett.20.1445}
  {\bibfield  {journal} {\bibinfo  {journal} {Physical Review Letters}\
  }\textbf {\bibinfo {volume} {20}},\ \bibinfo {pages} {1445} (\bibinfo {year}
  {1968})}\BibitemShut {NoStop}%
\bibitem [{\citenamefont {Kohn}(1964)}]{kohn1964theory}%
  \BibitemOpen
  \bibfield  {author} {\bibinfo {author} {\bibfnamefont {W.}~\bibnamefont
  {Kohn}},\ }\href {\doibase 10.1103/PhysRev.133.A171} {\bibfield  {journal}
  {\bibinfo  {journal} {Phys. Rev.}\ }\textbf {\bibinfo {volume} {133}},\
  \bibinfo {pages} {A171} (\bibinfo {year} {1964})}\BibitemShut {NoStop}%
\bibitem [{\citenamefont {Resta}(2002)}]{resta2002why}%
  \BibitemOpen
  \bibfield  {author} {\bibinfo {author} {\bibfnamefont {R.}~\bibnamefont
  {Resta}},\ }\href {\doibase 10.1088/0953-8984/14/20/201} {\bibfield
  {journal} {\bibinfo  {journal} {J. Phys.: Condens. Matter}\ }\textbf
  {\bibinfo {volume} {14}},\ \bibinfo {pages} {R625} (\bibinfo {year}
  {2002})}\BibitemShut {NoStop}%
\bibitem [{Note1()}]{Note1}%
  \BibitemOpen
  \bibinfo {note} {Here multiple factors have severely restricted evaluation of
  $D$ accurately for only small $N$: (i) use of periodic boundary conditions,
  (ii) then inapplicability of the MPO technique for approximating LR
  interactions, (iii) evaluation of the second order derivative in Eq.\ref
  {eq:stiff} being very numerical delicate.}\BibitemShut {Stop}%
\bibitem [{\citenamefont {Amico}\ \emph {et~al.}(2008)\citenamefont {Amico},
  \citenamefont {Fazio}, \citenamefont {Osterloh},\ and\ \citenamefont
  {Vedral}}]{amico2008entanglement}%
  \BibitemOpen
  \bibfield  {author} {\bibinfo {author} {\bibfnamefont {L.}~\bibnamefont
  {Amico}}, \bibinfo {author} {\bibfnamefont {R.}~\bibnamefont {Fazio}},
  \bibinfo {author} {\bibfnamefont {A.}~\bibnamefont {Osterloh}}, \ and\
  \bibinfo {author} {\bibfnamefont {V.}~\bibnamefont {Vedral}},\ }\href
  {\doibase 10.1103/RevModPhys.80.517} {\bibfield  {journal} {\bibinfo
  {journal} {Rev. Mod. Phys.}\ }\textbf {\bibinfo {volume} {80}},\ \bibinfo
  {pages} {517} (\bibinfo {year} {2008})}\BibitemShut {NoStop}%
\bibitem [{\citenamefont {Wang}\ \emph {et~al.}(2010)\citenamefont {Wang},
  \citenamefont {Feng},\ and\ \citenamefont {Chen}}]{wang2010berezinskii}%
  \BibitemOpen
  \bibfield  {author} {\bibinfo {author} {\bibfnamefont {B.}~\bibnamefont
  {Wang}}, \bibinfo {author} {\bibfnamefont {M.}~\bibnamefont {Feng}}, \ and\
  \bibinfo {author} {\bibfnamefont {Z.-Q.}\ \bibnamefont {Chen}},\ }\href
  {\doibase 10.1103/PhysRevA.81.064301} {\bibfield  {journal} {\bibinfo
  {journal} {Phys. Rev. A}\ }\textbf {\bibinfo {volume} {81}},\ \bibinfo
  {pages} {064301} (\bibinfo {year} {2010})}\BibitemShut {NoStop}%
\bibitem [{\citenamefont {Eisert}\ \emph {et~al.}(2010)\citenamefont {Eisert},
  \citenamefont {Cramer},\ and\ \citenamefont {Plenio}}]{eisert2010colloquium}%
  \BibitemOpen
  \bibfield  {author} {\bibinfo {author} {\bibfnamefont {J.}~\bibnamefont
  {Eisert}}, \bibinfo {author} {\bibfnamefont {M.}~\bibnamefont {Cramer}}, \
  and\ \bibinfo {author} {\bibfnamefont {M.~B.}\ \bibnamefont {Plenio}},\
  }\href {\doibase 10.1103/RevModPhys.82.277} {\bibfield  {journal} {\bibinfo
  {journal} {Rev. Mod. Phys.}\ }\textbf {\bibinfo {volume} {82}},\ \bibinfo
  {pages} {277} (\bibinfo {year} {2010})}\BibitemShut {NoStop}%
\bibitem [{\citenamefont {Hastings}(2007)}]{hastings2007area}%
  \BibitemOpen
  \bibfield  {author} {\bibinfo {author} {\bibfnamefont {M.~B.}\ \bibnamefont
  {Hastings}},\ }\href {\doibase 10.1088/1742-5468/2007/08/P08024} {\bibfield
  {journal} {\bibinfo  {journal} {J. Stat. Mech. Theor. Exp.}\ }\textbf
  {\bibinfo {volume} {2007}},\ \bibinfo {pages} {P08024} (\bibinfo {year}
  {2007})}\BibitemShut {NoStop}%
\bibitem [{\citenamefont {White}\ \emph {et~al.}(2002)\citenamefont {White},
  \citenamefont {Affleck},\ and\ \citenamefont {Scalapino}}]{white2002friedel}%
  \BibitemOpen
  \bibfield  {author} {\bibinfo {author} {\bibfnamefont {S.~R.}\ \bibnamefont
  {White}}, \bibinfo {author} {\bibfnamefont {I.}~\bibnamefont {Affleck}}, \
  and\ \bibinfo {author} {\bibfnamefont {D.~J.}\ \bibnamefont {Scalapino}},\
  }\href {\doibase 10.1103/PhysRevB.65.165122} {\bibfield  {journal} {\bibinfo
  {journal} {Phys. Rev. B}\ }\textbf {\bibinfo {volume} {65}},\ \bibinfo
  {pages} {165122} (\bibinfo {year} {2002})}\BibitemShut {NoStop}%
\bibitem [{\citenamefont {Wang}\ \emph {et~al.}(2001)\citenamefont {Wang},
  \citenamefont {Millis},\ and\ \citenamefont {Das~Sarma}}]{wang2001coulomb}%
  \BibitemOpen
  \bibfield  {author} {\bibinfo {author} {\bibfnamefont {D.~W.}\ \bibnamefont
  {Wang}}, \bibinfo {author} {\bibfnamefont {A.~J.}\ \bibnamefont {Millis}}, \
  and\ \bibinfo {author} {\bibfnamefont {S.}~\bibnamefont {Das~Sarma}},\ }\href
  {\doibase 10.1103/PhysRevB.64.193307} {\bibfield  {journal} {\bibinfo
  {journal} {Phys. Rev. B}\ }\textbf {\bibinfo {volume} {64}},\ \bibinfo
  {pages} {193307} (\bibinfo {year} {2001})}\BibitemShut {NoStop}%
\bibitem [{\citenamefont {Iucci}\ and\ \citenamefont
  {Na{\'o}n}(2000)}]{iucci2000exact}%
  \BibitemOpen
  \bibfield  {author} {\bibinfo {author} {\bibfnamefont {A.}~\bibnamefont
  {Iucci}}\ and\ \bibinfo {author} {\bibfnamefont {C.}~\bibnamefont
  {Na{\'o}n}},\ }\href {\doibase 10.1103/PhysRevB.61.15530} {\bibfield
  {journal} {\bibinfo  {journal} {Phys. Rev. B}\ }\textbf {\bibinfo {volume}
  {61}},\ \bibinfo {pages} {15530} (\bibinfo {year} {2000})}\BibitemShut
  {NoStop}%
\bibitem [{\citenamefont {Na{\'o}n}\ \emph {et~al.}(2005)\citenamefont
  {Na{\'o}n}, \citenamefont {Salvay},\ and\ \citenamefont
  {Trobo}}]{naon2005conformal}%
  \BibitemOpen
  \bibfield  {author} {\bibinfo {author} {\bibfnamefont {C.~M.}\ \bibnamefont
  {Na{\'o}n}}, \bibinfo {author} {\bibfnamefont {M.~J.}\ \bibnamefont
  {Salvay}}, \ and\ \bibinfo {author} {\bibfnamefont {M.~L.}\ \bibnamefont
  {Trobo}},\ }\href {\doibase 10.1103/PhysRevB.72.245110} {\bibfield  {journal}
  {\bibinfo  {journal} {Phys. Rev. B}\ }\textbf {\bibinfo {volume} {72}},\
  \bibinfo {pages} {245110} (\bibinfo {year} {2005})}\BibitemShut {NoStop}%
\bibitem [{\citenamefont {Inoue}\ and\ \citenamefont
  {Nomura}(2006)}]{inoue2006conformal}%
  \BibitemOpen
  \bibfield  {author} {\bibinfo {author} {\bibfnamefont {H.}~\bibnamefont
  {Inoue}}\ and\ \bibinfo {author} {\bibfnamefont {K.}~\bibnamefont {Nomura}},\
  }\href {\doibase 10.1088/0305-4470/39/9/012} {\bibfield  {journal} {\bibinfo
  {journal} {J. Phys. A: Math. Gen.}\ }\textbf {\bibinfo {volume} {39}},\
  \bibinfo {pages} {2161} (\bibinfo {year} {2006})}\BibitemShut {NoStop}%
\bibitem [{\citenamefont {Mishra}\ \emph {et~al.}(2011)\citenamefont {Mishra},
  \citenamefont {Carrasquilla},\ and\ \citenamefont {Rigol}}]{mishra2011phase}%
  \BibitemOpen
  \bibfield  {author} {\bibinfo {author} {\bibfnamefont {T.}~\bibnamefont
  {Mishra}}, \bibinfo {author} {\bibfnamefont {J.}~\bibnamefont
  {Carrasquilla}}, \ and\ \bibinfo {author} {\bibfnamefont {M.}~\bibnamefont
  {Rigol}},\ }\href {\doibase 10.1103/PhysRevB.84.115135} {\bibfield  {journal}
  {\bibinfo  {journal} {Phys. Rev. B}\ }\textbf {\bibinfo {volume} {84}},\
  \bibinfo {pages} {115135} (\bibinfo {year} {2011})}\BibitemShut {NoStop}%
\bibitem [{\citenamefont {Ejima}\ \emph
  {et~al.}(2005{\natexlab{b}})\citenamefont {Ejima}, \citenamefont {Gebhard},
  \citenamefont {Nishimoto},\ and\ \citenamefont {Ohta}}]{ejima2005phase}%
  \BibitemOpen
  \bibfield  {author} {\bibinfo {author} {\bibfnamefont {S.}~\bibnamefont
  {Ejima}}, \bibinfo {author} {\bibfnamefont {F.}~\bibnamefont {Gebhard}},
  \bibinfo {author} {\bibfnamefont {S.}~\bibnamefont {Nishimoto}}, \ and\
  \bibinfo {author} {\bibfnamefont {Y.}~\bibnamefont {Ohta}},\ }\href {\doibase
  10.1103/PhysRevB.72.033101} {\bibfield  {journal} {\bibinfo  {journal} {Phys.
  Rev. B}\ }\textbf {\bibinfo {volume} {72}},\ \bibinfo {pages} {033101}
  (\bibinfo {year} {2005}{\natexlab{b}})}\BibitemShut {NoStop}%
\end{thebibliography}%

 

\end{document}